\documentclass[11pt, a4paper, oneside]{article}

\pdfoutput=1

\usepackage[english]{babel}
\usepackage[utf8]{inputenc}
\usepackage{amsmath}
\usepackage{amssymb}
\usepackage{amsthm}
\usepackage{cite}
\usepackage{xcolor}
\usepackage[ruled,vlined]{algorithm2e}
\usepackage{float}
\usepackage{booktabs}
\usepackage{graphicx}
\usepackage{colortbl}
\usepackage[hidelinks]{hyperref}
\usepackage{adjustbox}
\usepackage{geometry}

\usepackage{natbib}
\SetKw{And}{and}
\SetKw{Break}{break}

\newcommand{\tfromi}{\tau_i} % truncation from this value
\newcommand{\tfrom}{\tau} % truncation from this value
\newcommand{\tto}{\gamma} % truncation to this value

\newcommand{\Pg}{\Pi} % group of transformations
\newcommand{\pib}{\boldsymbol{\pi}} % vector of random transformations
\newcommand{\qua}[1][V]{c_{#1}^{(\omega)}} % quantile for a set

\newcommand{\R}{\mathcal{R}} % sets rejeected by closed testing
\newcommand{\ds}{\delta(S)} % number of true discoveries
\newcommand{\dsa}{d(S)} % lower confidence bound for the number of true discoveries
\newcommand{\qsa}{q(S)} % upper confidence bound for the number of false discoveries

\newcommand{\pz}{\phi(z)} % indicator function
\newcommand{\underpz}{\underline{\phi}(z)} % indicator function
\newcommand{\overpz}{\overline{\phi}(z)} % indicator function
\newcommand{\underpzn}[1][n]{\underline{\phi}^{(#1)}(z)} % indicator function
\newcommand{\overpzn}[1][n]{\overline{\phi}^{(#1)}(z)} % indicator function

\newcommand{\V}{\mathcal{V}_z} % supersets of the subsets in \Z
\newcommand{\Vv}[1][v]{\mathcal{V}_z(#1)} % supersets of the subsets in \Z with size v
\newcommand{\low}[1][z]{u_{#1}} % lower bound
\newcommand{\up}[1][z]{\ell_{#1}} % lower bound
\newcommand{\lowv}[1][v]{u_z(#1)} % lower bound
\newcommand{\upv}[1][v]{\ell_z(#1)} % lower bound

\newcommand{\Vrem}[1][z]{\mathcal{V}_{#1}^{-}}
\newcommand{\Vkeep}[1][z]{\mathcal{V}_{#1}^{+}}

\newcommand{\hi}{\hat{i}}
\newcommand{\hj}{\hat{j}}

\newcommand{\ci}[1][1]{\theta_{#1}}

\newtheorem{theorem}{Theorem}

\newtheorem{lemma}{Lemma}
\newtheorem{prop}{Proposition}
\newtheorem{assumption}{Assumption}

\usepackage{authblk}
\title{Permutation-Based True Discovery Guarantee by Sum Tests}
\author[1]{Anna Vesely\footnote{\href{mailto:vesely@uni-bremen.de}{vesely@uni-bremen.de}}}
\author[2]{ Livio Finos\footnote{\href{mailto:livio.finos@unipd.it}{livio.finos@unipd.it}}}
\author[3]{Jelle J.~Goeman\footnote{\href{mailto:j.j.goeman@lumc.nl}{j.j.goeman@lumc.nl}}}
\affil[1]{\footnotesize Institute for Statistics, University of Bremen}
\affil[2]{\footnotesize Department of Statistical Sciences, University of Padua}
\affil[3]{\footnotesize Department of Biomedical Data Sciences, Leiden University Medical Center}
\date{}                     %% if you don't need date to appear
\date{}

\begin{document}
\maketitle

% --------------------------------------------------------------------------------------------------------------------------------------------------------------------------------------------------------------------------------------------------------------------------------------------------------------------
% --------------------------------------------------------------------------------------------------------------------------------------------------------------------------------------------------------------------------------------------------------------------------------------------------------------------

\begin{abstract}
Sum-based global tests are highly popular in multiple hypothesis testing. In this paper we propose a general closed testing procedure for sum tests, which provides lower confidence bounds for the proportion of true discoveries (TDP), simultaneously over all subsets of hypotheses. These simultaneous inferences come for free, i.e., without any adjustment of the $\alpha$-level, whenever a global test is used. Our method allows for an exploratory approach, as simultaneity ensures control of the TDP even when the subset of interest is selected post hoc. It adapts to the unknown joint distribution of the data through permutation testing. Any sum test may be employed, depending on the desired power properties. We present an iterative shortcut for the closed testing procedure, based on the branch and bound algorithm, which converges to the full closed testing results, often after few iterations; even if it is stopped early, it controls the TDP. We compare the properties of different choices for the sum test through simulations, then we illustrate the feasibility of the method for high dimensional data on brain imaging and genomics data.
\end{abstract}


% --------------------------------------------------------------------------------------------------------------------------------------------------------------------------------------------------------------------------------------------------------------------------------------------------------------------
% --------------------------------------------------------------------------------------------------------------------------------------------------------------------------------------------------------------------------------------------------------------------------------------------------------------------

\section{Introduction} \label{intro}
In high-dimensional data analysis, researchers are often interested in detecting subsets of features that are associated with a given outcome. For instance, in functional magnetic resonance imaging (fMRI) data the objective may be to identify a brain region that is activated by a stimulus; in genomics data one may want to find a biological pathway that is differentially expressed. In this context, global tests allow to aggregate signal from multiple features and make meaningful statements at the set level. A diverse range of global tests has been proposed in literature: well-known examples are p-value combinations, described and compared in \citet{pesarin}, \citet{comploughin}, \citet{compwon} and \citet{pesarinsalmaso}; other popular methods are Simes' test \citep{simes}, the global test of \citet{globalbis}, the sequence kernel association test (SKAT) \citep{skat} and higher criticism \citep{hc}. A substantial proportion, including many of the above-mentioned methods, is sum-based, meaning that the global test statistic may be written as a sum of contributions per feature. In this paper we restrict to such sum-based tests.

The probability distribution of a global statistic depends not only on the marginal distributions of the data, but also on the joint distribution; for this reason, many sum tests only have a known null distribution under independence. Approaches that deal with the a-priori unknown joint distribution are worst-case distributions, defined either generally or under restrictive assumptions \citep{paverage}, and nonparametric permutation testing \citep{perm1, perm2}. As worst-case distributions tend to be very conservative, the latter approach is preferable; it relies on minimal assumptions \citep{exact}, and generally offers an improvement in power over the parametric approach, especially when multiple hypotheses are considered \citep{westyoung, pesarin, exact0, fdpbounds}.

Rejecting a null hypothesis, however, gives little information on the corresponding set. A significant p-value only indicates that there is at least one true discovery, i.e., one feature associated with the outcome, but does not give any information on the proportion of true discoveries (TDP), nor their localization. This becomes problematic especially for large sets \citep{woo}. Moreover, since interest is usually not just in the set of all features, but in several subsets, a multiple testing procedure is necessary \citep{multiplefmri, multiplegene}. Finally, when researchers do not know a priori which subsets they are interested in, they may want to test many and then make the selection post hoc. The case for the use of TDPs in large-scale testing problems was argued by \citet{ari} in neuroimaging and by \citet{sea} in genomics.

This paper presents a general approach for inference on the TDP. The method allows any sum-based test, requiring only that critical values are determined by permutations. It provides TDPs not only for the full testing problem, but also simultaneously for all subsets, allowing subsets of interest to be chosen post hoc.

We will rely on the closed testing framework \citep{closed}, which allows to construct confidence sets for the TDP simultaneously over all possible subsets \citep{genovese2, exploratory, simultaneous}. These additional simultaneous inferences on all subsets come for free, i.e., without any adjustment of the $\alpha$-level, whenever a global test is applied. Simultaneity ensures that the procedure is not compromised by post-hoc selection, therefore researchers can postpone the choice of the subset until after seeing the data, while still obtaining valid confidence sets; used in this way, closed testing allows a form of post-hoc inference. Furthermore, closed testing has been proven to be the optimal way to construct multiple testing procedures, as all family-wise error rate (FWER), TDP and related methods are either equivalent to or can be improved by it \citep{only}. The main challenge is the computational complexity, which is extremely high when considering many hypotheses, and when using many permutations. Permutation-based closed testing for the TDP so far mostly focused on Simes-based test procedures, while sum tests were approached under independence or with worst-case distributions \citep{paverage, harmonic, largetdp}, that are simpler as critical values depend only on the size of the subset.

We propose a general closed testing procedure for sum-based permutation tests, which provides simultaneous confidence sets for the TDP of all subsets of the testing problem. We develop two shortcuts to make this procedure feasible for large-scale problems. First, we develop a quick shortcut that approximates closed testing and has worst-case complexity of order $m\log^2 m$ in the number $m$ of individual hypotheses, and linearithmic in the number of permutations. Next, we embed this shortcut within a branch and bound algorithm, obtaining an iterative procedure that converges to full closed testing, often after few iterations; even if it is stopped early, it still controls the TDP. This procedure is exact and extremely flexible, as it applies to any sum test and adapts to the correlation structure of the data. It can be scaled up to high-dimensional problems, such as fMRI data, whose typical dimension is of order $10^5$. Finally, we show that particular choices of the sum test statistic, namely statistics based on truncation, result in faster procedures.

The structure of the paper is as follows. First, we briefly discuss related works in Section \ref{relworks}. Then we introduce sum tests in Section \ref{sums}, and we review the properties of permutation testing and closed testing in Sections \ref{perm} and \ref{closed}. We derive the single-step shortcut in Section \ref{shortcut}, and characterize when it is equivalent to closed testing in Section \ref{equivalence}. In Section \ref{bab} we define the iterative shortcut, and finally in Section \ref{truncation} we introduce refinements that improve the computational complexity. In the remaining section we compare the properties of different sum tests through simulations, and explore an application to fMRI data. Proofs and some additional results are postponed to the appendix.

% --------------------------------------------------------------------------------------------------------------------------------------------------------------------------------------------------------------------------------------------------------------------------------------------------------------------
% --------------------------------------------------------------------------------------------------------------------------------------------------------------------------------------------------------------------------------------------------------------------------------------------------------------------

\section{Related work} \label{relworks}
In this section we discuss related work, highlighting the contribution of the proposed method and its relevance in applications. As argued in Section \ref{intro}, in this paper we focus on permutation-based tests. Here we justify the choice of closed testing procedures that give lower ($1-\alpha$)-confidence bounds for the TDP simultaneously over all subsets of hypotheses, which we will refer to as procedures with true discovery guarantee as in \citet{only}. Then we argue that it is worthwhile to construct such procedures for global tests that are frequently used, many of which are sum-based.

\citet{genovese2} and \citet{exploratory} showed that all global tests automatically come with an inbuilt selective inference method; they can be embedded in the closed testing framework to obtain procedures with true discovery guarantee without any adjustment of the $\alpha$-level. Furthermore, a great number of multiple testing methods, including all those controlling FWER, generalized FWER ($k$-FWER), false discovery proportion (FDP), false discovery exceedance (FDX) and joint error rate (JER), can be written as procedures with true discovery guarantee. Among these, however, only closed testing procedures are admissible, i.e., cannot be uniformly improved \citep{only}. This motivates the study of closed testing procedures for popular global tests.

So far, most procedures that explicitly give true discovery guarantee \citep{meinshausen,ari,fdpbounds,sea,sanssouci,pARI,notip} were constructed using critical vectors for ordered p-values, e.g., based on variants of \citet{simes} or higher criticism \citep{hc}. With the exception of higher criticism, the global tests implicit in these procedures have seldom been considered as global tests in application contexts, and their popularity in multiple testing procedures is partly motivated by mathematical convenience. In contrast, tests based on sums are natural and popular as global tests. This broad class includes many popular p-value combination tests, such as the classical Fisher combination \citep{fisher}, as well as recent proposals such as \citet{harmonic}, \citet{cauchy}, the global test of \citet{globalbis}, SKAT \citep{skat}, and e-value combinations \citep{evalues}. Though closed testing procedures for sum-based tests were proposed in general in the parametric approach \citep{largetdp} and for some particular cases \citep{exploratory,sanssouci}, general scalable procedures in the permutation framework were lacking. In this paper we fill this gap, providing a procedure that can be applied to any sum-based test, as long as permutations are used to calculate the critical values.

Among permutation-based procedures, we mention especially the methods of \citet{sanssouci} and \citet{pARI}, using tests based on critical vectors of ordered p-values. First, we remark that our proposed method is not a competitor but complementary, as it deals with a different choice of the underlying test with different power properties. Subsequently, we observe that these methods do not perform full closed testing, and thus may be conservative. \citet{sanssouci} and the single-step version in \citet{pARI} have computation times primarily related to computing and sorting permutation test statistics; we will show that the computation time of our single-step shortcut is comparable. The iterative method of \citet{pARI} uniformly improves the corresponding single-step version and \citet{sanssouci}, but requires a high computational time and is still not guaranteed to converge to closed testing. On the contrary, the proposed iterative shortcut converges to closed testing and so cannot be uniformly improved.

%We conclude by mentioning two other related methods. First, the extension of significance analysis of microarrays (SAM) of \citet{exact0}, that gives confidence bounds for the FDP for a pre-specified rejection region, is a special case of both \citet{fdpbounds} and the proposed method. Secondly, the work of \citet{largetdp} is complementary to this paper, as it implements an analogous approach in the parametric framework. In Section \ref{sims} we will compare the two in the case of generalized means of p-values \citep{paverage}.

% --------------------------------------------------------------------------------------------------------------------------------------------------------------------------------------------------------------------------------------------------------------------------------------------------------------------
% --------------------------------------------------------------------------------------------------------------------------------------------------------------------------------------------------------------------------------------------------------------------------------------------------------------------

\section{Sum tests} \label{sums}
We start with a general definition of a sum test statistic. Throughout the paper, we will refer to null hypotheses simply as hypotheses, and we will denote both variables and sets with capital letters, leaving the distinction to context. Let $\mathbf{X}=(X_1,\ldots,X_m)$ be a collection of observable variables from $m$ testing units, having indices in $M=\{1,\ldots,m\}$ and taking values in a sample space $\mathcal{X}$. We are interested in studying $m$ corresponding univariate hypotheses $H_1,\ldots, H_m$ with confidence $1-\alpha$, where $\alpha\in[0,1)$. \ Let $N\subseteq M$ be the unknown subset of true hypotheses. A generic subset $S\subseteq M$, with size $|S|=s$, defines an intersection hypothesis $H_S=\bigcap_{i\in S}H_i$, which is true if and only if $S\subseteq N$. In the particular case of $S=\emptyset$, we take $H_\emptyset$ as usual to be a hypothesis that is always true.

For each univariate hypothesis $H_i$, let $T_i:\mathcal{X}\rightarrow\mathbb{R}$ be a test statistic. The general form of a sum test statistic for $H_S$ is
\[T_S = g\left(\sum_{i\in S}f_i(T_i)\right),\]
where $f_i:\mathbb{R}\rightarrow\mathbb{R}$ are generic functions, and $g:\mathbb{R}\rightarrow\mathbb{R}$ is strictly monotone. Usually the functions $f_i$ are also taken as monotone, so that high values of $T_S$ give evidence against $H_S$. Moreover, as $f_i$ may depend on $i$, the contributions $f_i(T_i)$ may have different distributions, as in the case of weighted sums. Examples include p-value combinations such as \citet{fisher}, \citet{pearson}, Liptak/Stouffer \citep{liptak}, \citet{lancaster}, \citet{edgington}, and Cauchy \citep{cauchy}. We mention especially the generalized mean family \citep{paverage} with $f_i(y)=y^r$ and $g(z)=z^{1/r}$, where $r\in\mathbb{R}$, for which \citet{harmonic} studied the harmonic mean ($r=-1$).

Since we can always re-write $\tilde{T}_i=f_i(T_i)$ and $\tilde{T}_S=g^{-1}(T_S),$ without loss of generality we can assume that $f_i$ and $g$ are the identity, so that $T_S = \sum_{i\in S}T_i$. In particular, for the empty set we obtain $T_{\emptyset}=0$. Furthermore, we assume that the signs of the statistics $T_i$ are chosen in such a way that high values of $T_i$, and therefore $T_S$, correspond to evidence against $H_i$ and $H_S$, respectively.

% --------------------------------------------------------------------------------------------------------------------------------------------------------------------------------------------------------------------------------------------------------------------------------------------------------------------
% --------------------------------------------------------------------------------------------------------------------------------------------------------------------------------------------------------------------------------------------------------------------------------------------------------------------

\section{Permutation testing} \label{perm}
To test $H_S$ with significance level $\alpha$ we will use permutations. Let $\Pg$ be a collection of transformations $\pi :\mathcal{X}\rightarrow\mathcal{X}$ of the sample space; these may be permutations, but also other transformations such as rotations \citep{rotat,rotations} and sign flipping \citep{score}. We assume that $\Pg$ is an algebraic group with respect to the operation of composition of functions. The group structure is important as, without it, the resulting test may be highly conservative or anti-conservative \citep{hoeff, south}.

Denote with $T_i=T_i(\mathbf{X})$ and $T_i^{\pi}=T_i(\pi \mathbf{X})$, with $\pi\in\Pg$, the statistics for the original and transformed variables, respectively, and with $t_i$ and $t_i^\pi$ the values computed on the observed and transformed data. The main assumption of permutation testing is the following.

\begin{assumption}\label{A:perm}
The joint distribution of the statistics $T_i^{\pi}$, with $i\in N$ and $\pi\in\Pg$, is invariant under all transformations in $\Pg$ of $\mathbf{X}$: $(T_i)_{i\in N}\overset{\text{d}}{=} (T_i^\pi)_{i\in N}$ for each $\pi\in\Pg$, where $\overset{\text{d}}{=}$ denotes equality in distribution.
\end{assumption}

This assumption is common to most permutation-based multiple-testing methods, such as maxT-method \citep{westyoung, meinshausen, prclosed, fdpbounds}. For some choices of the group $\Pg$, the assumption holds only asymptotically \citep{permglm, rotations, score}. Detailed illustration and examples can be found in \citet{pesarin}, \citet{toperm} and \citet{exact}. Even if the invariance assumption is common and reasonable in many contexts, in applications an argument must be given for it; in some cases, it is violated even asymptotically (e.g., for Behrens-Fisher problem \citep{schild}).

A slightly stronger assumption, that is easier to check, is that the statistic $T_S=T_S(\mathbf{X}_S)$ is a function of $\mathbf{X}_S=(X_i\,:\,i\in S)$ only, and $\mathbf{X}_N\overset{\text{d}}{=} \pi \mathbf{X}_N$ for each $\pi$. Note that the assumption holds also when the distributions of the individual statistics $T_i$ are different, as in the case of weighted sums. Moreover, it holds in the particular case when $H_S$ true implies that $\mathbf{X}_S\overset{\text{d}}{=} \pi \mathbf{X}_S$ for each $\pi$.

If the cardinality of $\Pg$ is large, a valid $\alpha$-level test may use $B$ randomly chosen elements \citep{exact0}. The value of $B$ does not need to grow with $m$ or $s$; to have non-zero power we must only have $B\geq 1/\alpha$, though larger values of $B$ give more power. For $\alpha=0.05$, $B\geq 200$ is generally sufficient (see Section \ref{appl:fmri}).
%Moreover, when $B$ is a multiple of $1/\alpha$ the test is exact \citep{exact}.
Consider a vector $\pib=(\pi_1,\ldots,\pi_B)$, where $\pi_1=\text{id}$ is the identity in $\Pg$, and $\pi_2,\ldots,\pi_B$ are random elements drawn with replacement from a uniform distribution on $\Pg$. A test for $H_S$ may be defined taking as critical value the $\lceil (1-\alpha) B\rceil$-th quantile, where $\lceil\cdot\rceil$ represents the ceiling function, and $t_S^{(1)}\leq\ldots\leq t_S^{(B)}$ are the sorted values $t_S^\pi$, with $\pi\in\pib$.

\begin{lemma}\label{L:perm}
Under Assumption \ref{A:perm}, the test that rejects $H_S$ when $t_S > t_S^{(\lceil (1-\alpha) B\rceil)}$ is an $\alpha$-level test.
\end{lemma}

The test is defined conditionally on $\mathbf{X}$, but it becomes unconditional if we take the expected value on both sides of the inequality. Note that both the test statistic and the critical value are random variables. For our method it will be convenient to use an equivalent characterization of the test with a non-random critical value. Therefore, for each $\pi$ we define the centered statistic $C_S^{\pi}=  T_S - T_S^{\pi}$, so that the observed value $c_S=c_S^{\text{id}}$ is always zero, and so no longer random. We give a permutation test based on these new statistics, using $\omega=\lfloor\alpha B\rfloor +1$ to obtain the quantile, where $\lfloor\cdot\rfloor$ is the floor function.

\begin{theorem}\label{T:perm}
Under Assumption \ref{A:perm}, the test that rejects $H_S$ when $c_S^{(\lfloor\alpha B\rfloor +1)}>0$ is an $\alpha$-level test.
\end{theorem}

For illustration, we introduce a recurring toy example with $m=5$ univariate hypotheses and $B=6$ transformations (Table \ref{Table:toystart}). Given the subset $S=\{1,2\}$, we are interested in testing $H_S$ with significance level $\alpha=0.4$. The statistics $t_S^\pi$ and $c_S^\pi$ are obtained summing columns 1 and 2 by row. Since $\omega=3$ and $c_S^{(\omega)}=2$, the test of Theorem \ref{T:perm} rejects $H_S$.

\begin{table}
\caption{\label{Table:toystart} Toy example: original and centered test statistics.}
\centering
\begin{tabular}{ccccccccccccc}
\toprule
 &  & \multicolumn{5}{c}{original $t_i^\pi$} &  &  \multicolumn{5}{c}{centered $c_i^\pi$}\\
 &  & $H_1$ & $H_2$ & $H_3$ & $H_4$ & $H_5$ &  & $H_1$ & $H_2$ & $H_3$ & $H_4$ & $H_5$\\
\midrule
id &  & 6 & 5 & 4 & 1 & 1 &  & 0 & 0 & 0 & 0 & 0\\
$\pi_2$ &  & 1 & 2 & 1 & 0 & 4 &  & 5 & 3 & 3 & 1 & -3\\
$\pi_3$ &  & 8 & 3 & 0 & 2 & 1 &  & -2 & 2 & 4 & -1 & 0\\
$\pi_4$ &  & 8 & 1 & 0 & 1 & 0 &  & -2 & 4 & 4 & 0 & 1\\
$\pi_5$ &  & 0 & 6 & 1 & 1 & 2 &  & 6 & -1 & 3 & 0 & -1\\
$\pi_6$ &  & 7 & 0 & 1 & 2 & 1 &  & -1 & 5 & 3 & -1 & 0\\
\bottomrule
\end{tabular}
\end{table}

% --------------------------------------------------------------------------------------------------------------------------------------------------------------------------------------------------------------------------------------------------------------------------------------------------------------------
% --------------------------------------------------------------------------------------------------------------------------------------------------------------------------------------------------------------------------------------------------------------------------------------------------------------------

\section{True discovery guarantee} \label{closed}
Based on the notation introduced above, consider the number of true discoveries $\ds=|S\setminus N|$ made when rejecting $H_S$. We are interested in deriving simultaneous ($1-\alpha$)-confidence sets for this number, so that the simultaneity makes their coverage robust against post-hoc selection. This way, the rejected hypothesis can be selected after reviewing all confidence sets, while still keeping correct ($1 -\alpha$)-coverage of the corresponding confidence set \citep{exploratory}.

Let $d:2^M\rightarrow\mathbb{R}$ be a random function, where $2^M$ is the power set of $M$. We say that $d$ has true discovery guarantee if $\dsa$ are simultaneous lower ($1-\alpha$)-confidence bounds for $\ds$, i.e.,
\[P\left(\ds\geq\dsa\;\text{for each } S\subseteq M\right)\geq 1-\alpha .\]
An equivalent condition is that $\{d(S),\ldots,s\}$ is a ($1-\alpha$)-confidence set for $\ds$, simultaneously for all $S\subseteq M$. Notice that the resulting confidence sets are one-sided, since hypothesis testing is focused on rejecting, not accepting. From $d(S)$ simultaneous ($1-\alpha$)-confidence sets can be immediately derived for other quantities of interest such as the TDP and the number or proportion of false discoveries \citep{exploratory}.

A general way to construct procedures with true discovery guarantee is provided by closed testing, based on the principle of testing different subsets by means of a valid $\alpha$-level local test, which in this case is the permutation test. Throughout this paper, we will loosely say that a set $S$ is rejected when the corresponding hypothesis $H_S$ is. Hence denote the collection of sets rejected by the permutation test of Theorem \ref{T:perm} by 
\[\R = \left\{S\subseteq M\,:\, \qua[S]>0\right\}.\]
\citet{genovese2} and \citet{exploratory} equivalently define a procedure $d$ with true discovery guarantee as $\dsa=s - \qsa$, where
\begin{align}
\qsa=\max\left\{|V\cap S|\,:\,V\subseteq M,\,V\notin\R\right\} \label{def:q_first}
\end{align}
is the maximum intersection between $S$ and a set not rejected by the permutation test. The equivalence of the two methods is shown in \citet{only}.

The main challenge of this method is its exponential complexity in the number of hypotheses. Indeed, the number of tests that must be evaluated to determine $\dsa$ may be up to order $2^m$. In the toy example, where $m=5$, this number is $32$; it is immediate that it quickly grows to an infeasible size as $m$ increases.

%In order to appreciate the scale of the problem, consider the toy example, where $m=5$; computing $\dsa$ may require up to $32$ tests. It is immediate that this number quickly grows to an infeasible size as the number of hypotheses increases.

% --------------------------------------------------------------------------------------------------------------------------------------------------------------------------------------------------------------------------------------------------------------------------------------------------------------------
% --------------------------------------------------------------------------------------------------------------------------------------------------------------------------------------------------------------------------------------------------------------------------------------------------------------------

\section{Shortcut} \label{shortcut}
Fix the set of interest $S$, so that any dependence on it may be omitted in the notation. We propose a shortcut that quickly evaluates whether $q<z$ for any value $z$. This will allow to approximate $q$, and eventually define a procedure with true discovery guarantee. First, we will re-write $q$ as the unique change-point of an increasing function:
\begin{align}
\phi:\{0,\ldots,s+1\}\longrightarrow\{0,1\},\qquad\pz=1\quad\text{if and only if}\quad q<z  \label{prop:pz}\\
q=\max\left\{z\in\{0,\ldots,s+1\}\,:\,\pz=0\right\}. \label{def:q}
\end{align}
Then we will approximate $q$ from above with the change point $q^{(0)}$ of a second increasing function:
\begin{align}
\underline{\phi}:\{0,\ldots,s+1\}\longrightarrow\{0,1\}, \qquad\underpz \leq \pz\label{prop:underpz}\\
q^{(0)}=\max\left\{z\in\{0,\ldots,s+1\}\,:\,\underpz =0\right\}. \label{def:q0}
\end{align}

%In this Section we construct the functions $\phi$ and $\underline{\phi}$ so that they have the desired properties. Subsequently, this shortcut will be refined in different ways. In Section \ref{equivalence}, we discuss a sufficient condition for establishing that the shortcut is equivalent to closed testing for some $z$, i.e., $\underpz=\pz$. Next, in Section \ref{bab} we define an iterative shortcut which increases the power and eventually converges to the full closed testing result. Finally, in Section \ref{truncation} we argue that particular choices of the test statistic result in faster procedures.

% --------------------------------------------------------------------------------------------------------------------------------------------------------------------------------------------------------------------------------------------------------------------------------------------------------------------

%\subsection{Structure of the Shortcut}\label{shortstr}

We start by giving an equivalent characterization of the quantity of interest $q$. For any $z\in\{0,\ldots,s+1\}$, we define the collection $\V=\{V\subseteq M\,:\,|V\cap S|\geq z\}$ of sets that have at least size $z$ overlap with $S$, and investigate whether all its elements are rejected. We define $\phi$ so that it represents such rejection, taking
\begin{align}
\pz=\mathbf{1}\{\V\subseteq\R\}\qquad (z\in\{0,\ldots,s+1\}), \label{def:pz}
\end{align}
where $\mathbf{1}\{\cdot\}$ denotes the indicator function. The following lemma shows that $q$ can be written as in \eqref{def:q}.

\begin{lemma}\label{L:phi_q}
$\phi(0)=0$ and $\phi(s+1)=1$. Moreover, $\pz=0$ if and only if $z\in\{0,\ldots,q\}$.
\end{lemma}

Now we fix a value $z\in\{1,\ldots,s\}$ and derive the shortcut to make statements on $\pz$ without testing all the sets contained in $\V$. We do this by partitioning $\V$ by the size of its elements, obtaining
\begin{align}
\V=\bigcap_{v=z}^m\Vv,\qquad\Vv=\{V\in\V\,:\,|V|=v\}. \label{def:partition}
\end{align}
Each $\Vv$ is the sub-collection of all sets of size $v$ that have at least size $z$ overlap with $S$. We can analyse these sub-collections separately and combine the results, noting that $\pz=1$ if and only if $\Vv\subseteq\R$ for all $v\in\{z,\ldots,m\}$.

\begin{figure}
\centering
\makebox{\includegraphics[width=0.9\textwidth]{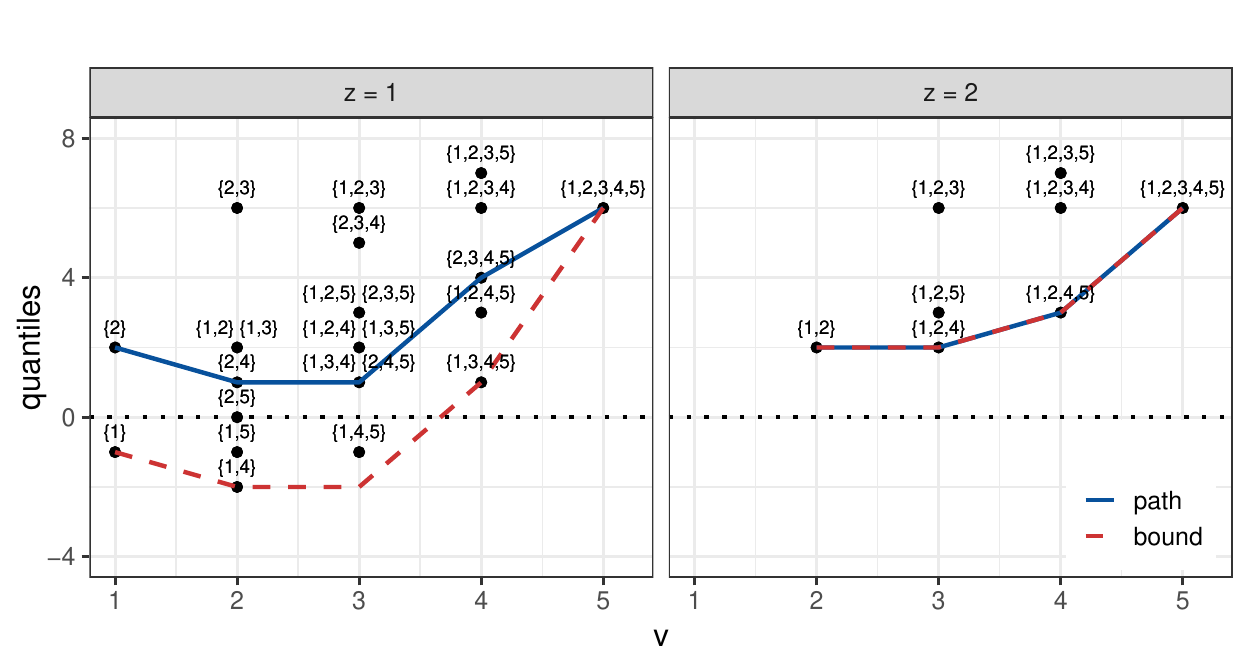}}
\caption{\label{Plot:base2} Toy example with $S=\{1,2\}$: shortcut to evaluate $\phi(z)$ in $z=1$ and $z=2$. Points denote the quantiles for the sets in $\mathcal{V}_z$. The dashed and solid lines represent the bound $\ell_z$ \eqref{prop:Uz} and the path $u_z$ \eqref{prop:Lz}, respectively.}
\end{figure}

By definition, $\Vv\subseteq\R$ when all sets in the sub-collection have positive quantiles, i.e., $\qua>0$ for each $V\in\Vv$. The main idea of the shortcut is to obtain information on each sub-collection $\Vv$ by bounding the corresponding quantiles from below. In particular, we will construct a bound
\begin{align}
\up\,:\,\{z,\ldots,m\}\longrightarrow\mathbb{R},\qquad \upv \leq \qua\quad\text{for each } V\in\Vv. \label{prop:Uz}
\end{align}
This way, if $\upv >0$, we know that all sets in $\Vv$ have positive quantiles. If $\up$ is positive in its entire domain, then $\Vv\subseteq\R$ for each $v$, and so $\pz=1$. Figure \ref{Plot:base2} displays the bound, which we will define in the following paragraphs, in the toy example for $z=1$ and $z=2$. Note that indeed all quantiles lie on it or above; the bound can be loose, as seen with $\up[1](3)$. Since $\up[2]$ lies entirely in the positive half-space, we know that $\phi(2)=1$. In contrast, we cannot make a statement on $\phi(1)$ based on $\up[1]$.

% --------------------------------------------------------------------------------------------------------------------------------------------------------------------------------------------------------------------------------------------------------------------------------------------------------------------

%\subsection{Bound}\label{upperb}

Fix a size $v\in\{z,\ldots,m\}$. To define an $\upv$ that does not exceed the minimum quantile over all sets in $\Vv$, as required in \eqref{prop:Uz}, we approximate the minimum quantile from below with the quantile of the minimum. We do this by taking the smallest centered statistics for each transformation $\pi$, with some constraints from the structure of $\Vv$.

In the toy example, choose $z=1$, and let $V$ be any set in the sub-collection $\mathcal{V}_1(v)$ of interest. Note that $V$ must contain $v$ indices, at least $z=1$ of which is in $S$. Consider the centered statistics $c_i^{\pi_2}$ for transformation $\pi_2$ (second row in Table \ref{Table:toystart}, right). First, we select the lowest value in $S$, then we sort the remaining values in ascending order, as in the second row of Table \ref{Table:toyup}. If $b_v^{\pi_2}$ is the sum of the first $v$ elements of the row, we know that $b_v^{\pi_2}\leq c_V^{\pi_2}$. After constructing the other rows of Table \ref{Table:toyup} according to the same principle, we define $\up[1](v)=b_v^{(\omega)}$. Since $b_v^\pi\leq c_V^{\pi}$ for each $\pi$, we obtain $\up[1](v)\leq\qua$.

\begin{table}
\caption{\label{Table:toyup} Toy example with $S=\{1,2\}$: matrix of the sorted centered statistics to compute the bound $\ell_1$. The value $\ell_1(v)$ is obtained summing the first $v$ columns by row, and then taking the quantile.}
\centering
\begin{tabular}{cccccccc}
\toprule
 &  & \multicolumn{1}{c}{selected in $S$} &  &  \multicolumn{4}{c}{remaining}\\
 &  & $i_1(\pi)$ &  & $j_1(\pi)$ & $j_2(\pi)$ & $j_3(\pi)$ & $j_4(\pi)$\\
\midrule
id &  & 0 ($H_1$) &  & 0 ($H_2$) & 0 ($H_3$) & 0 ($H_4$) & 0 ($H_5$)\\
$\pi_2$ &  & 3 ($H_2$) &  & -3 ($H_5$) & 1 ($H_4$) & 3 ($H_3$) & 5 ($H_1$)\\
$\pi_3$ &  & -2 ($H_1$) &  & -1 ($H_4$) & 0 ($H_5$) & 2 ($H_2$) & 4 ($H_3$)\\
$\pi_4$ &  & -2 ($H_1$) &  & 0 ($H_4$) & 1 ($H_5$) & 4 ($H_2$) & 4 ($H_3$)\\
$\pi_5$ &  & -1 ($H_2$) &  & -1 ($H_5$) & 0 ($H_4$) & 3 ($H_3$) &  6 ($H_1$)\\
$\pi_6$ &  & -1 ($H_1$) &  & -1 ($H_4$) & 0 ($H_5$) & 3 ($H_3$) & 5 ($H_2$)\\
\bottomrule
\end{tabular}
\end{table}

In general, for each $\pi\in\pib$, we select the $z$ smallest centered statistics in $S$, and then the $v-z$ remaining smallest statistics. We define two permutations of the indices:
\begin{align}
S=\{i_1(\pi),\ldots,i_s(\pi)\}\quad :\quad &c_{i_1(\pi)}^{\pi}\leq\ldots\leq c_{i_s(\pi)}^{\pi}  \label{permbound1}\\
M\setminus \{i_1(\pi),\ldots,i_z(\pi)\}
=\{j_1(\pi),\ldots,j_{m-z}(\pi)\}
\quad :\quad &c_{j_1(\pi)}^{\pi}\leq\ldots\leq c_{j_{m-z}(\pi)}^{\pi}.  \label{permbound2}
\end{align}
The set $\{i_1(\pi),\ldots,i_z(\pi)\}$ is a subset of $S$, containing the indices of the $z$ smallest values in $S$ (for transformation $\pi$). For instance, in the toy example we have $S=\{2,1\}$, and $M\setminus\{2\}=\{5,4,3,1\}$. Then the value of the bound is defined as
\begin{align}
\upv=b_v^{(\omega)}\qquad\text{where}\qquad b_v^{\pi} = \sum_{h=1}^{z} c_{i_h}^{\pi} + \sum_{h=1}^{v-z} c_{j_h}^{\pi}\quad (\pi\in\pib). \label{def:Uz}
\end{align}

\begin{lemma}\label{L:upper}
$\upv\leq\qua$ for all $V\in\Vv$. Hence $\min_v\upv >0$ implies $\pz=1$.
\end{lemma}

Now we use the bound to define a function $\underline{\phi}$ as in \eqref{prop:underpz}. In the extremes, where the value of $\phi$ is known, we set $\underline{\phi}(0)=\phi(0)=0$ and $\underline{\phi}(s+1)=\phi(s+1)=1$ (see Lemma \ref{L:phi_q}). Elsewhere, we set
\begin{align}
\underpz=\mathbf{1}\left\{\min_{v}\upv > 0\right\}\qquad (z\in\{1,\ldots,s\}). \label{def:upz}
\end{align}
This function may not be monotonic, but we are only interested in its smallest change point; indeed, if $\underpz=1$ for a value $z$, we know that $q<z$. We make it increasing and obtain a single change point in $q^{(0)}$, as defined in \eqref{def:q0}, by imposing
\begin{align}
\underpz=1\quad\text{if}\quad \underline{\phi}(z^*)=1 \text{ for some } z^*\leq z \qquad (z\in\{1,\ldots,s\}). \label{def:upz_bis}
\end{align}

\begin{prop}\label{P:short}
As $\underpz\leq\pz$ for each $z\in\{0,\ldots,s+1\}$, $q^{(0)}\geq q$.
\end{prop}

For instance, in the toy example of Figure \ref{Plot:base2}, $\underline{\phi}(1)=0$ and $\underline{\phi}(2)=1$, and so $q^{(0)}=1$. Finally, from this result we can approximate $d$ from below with $d^{(0)}=s-q^{(0)}$.

\begin{theorem}\label{T:d0}
$d^{(0)}\leq d$.
\end{theorem}

To summarise, Proposition \ref{P:short} represents the basis of the shortcut. For any value $z$, it allows to make statements on the value of $\pz$ by constructing $\underpz\leq\pz$; it requires to evaluate a number of tests which is linear in the total number $m$ of hypotheses, in contrast to the exponential number required by closed testing. Theorem \ref{T:d0} employs the shortcut to provide a lower ($1-\alpha$)-confidence bound $d^{(0)}$ for the number of true discoveries $\delta$. The theorem holds for all $S \subseteq M$, hence the procedure $d^{(0)}$ has true discovery guarantee. In Appendix \ref{alg} we propose an algorithm for the shortcut, then we embed it into a binary search to approximate $q$ with reduced complexity. We prove that in the worst case the computational complexity is of order $mB(\log^2 m +\log B)$. Moreover, we show how the method can be combined with an algorithm of \citet{largetdp} to find the largest set with given TDP among a collection of incremental sets.

% --------------------------------------------------------------------------------------------------------------------------------------------------------------------------------------------------------------------------------------------------------------------------------------------------------------------
% --------------------------------------------------------------------------------------------------------------------------------------------------------------------------------------------------------------------------------------------------------------------------------------------------------------------

\section{Equivalence to closed testing} \label{equivalence}
The shortcut of Proposition \ref{P:short} defines $\underpz\leq\pz$ for any $z$. For those values of $z$ for which $\underpz =1$, we know that also $\pz=1$. Where $\underpz = 0$, however, there are two distinct cases. If $\pz=0$, the shortcut is equivalent to closed testing; otherwise, if $\pz=1$, it is conservative, as it does not reject all sets in $\V$ while closed testing does. In the toy example with $z=1$ we are in the first case (Figure \ref{Plot:base2}, left), but we cannot see that from the bound only. Now we propose a sufficient condition to state that $\underpz=\pz$. This will play an important role in the iterative shortcut of Section \ref{bab}. We will define an increasing function
\begin{align}
\overline{\phi}:\{0,\ldots,s+1\}\longrightarrow\{0,1\},\qquad\underpz\leq\pz\leq\overpz . \label{prop:overpz}
\end{align}
This way, if $\underpz=\overpz$ for a value $z$, we know that $\underpz=\pz$. Note that this holds in particular when either $\underpz=1$ or $\overpz=0$.

Fix $z\in\{1,\ldots,s\}$. Based on partition \eqref{def:partition} of $\V$, the main idea is to construct a greedy path of sets $V_z \subset \ldots \subset V_m$, with $V_v \in \mathcal{V}_z(v)$ for each $v$, and check whether their quantiles are all strictly positive. If we find a non-positive quantile, then we have established that $\V \not\subseteq \R$, and so $\underpz=\pz=0$; the shortcut is equivalent to closed testing for this value of $z$. We will define the path
\begin{align}
\low\,:\,\{z,\ldots,m\}\longrightarrow\mathbb{R},\qquad\lowv = \qua[V_v]\quad\text{with}\quad V_v\in\Vv \label{prop:Lz}
\end{align}
that connects these quantiles. This way, if $\lowv\leq 0$, we know that $\Vv$ contains a non-rejected set, and so $\pz=0$. Figure \ref{Plot:base2} displays the bound $\up$ and the path $\low$, which we will define in the next paragraphs, for the toy example. The path connects some of the quantiles, one for each size $v$, and so is never smaller than the bound. From $\up[2]$ we already had $\phi(2)=1$; as $\low[1]$ is entirely positive, results on $\phi(1)$ are still unsure.

% -------------------------------------------------------#----------------------------------------------------------------------------------------------------------------------------------------------------------------------------------------------------------------------------------------------------------

%\subsection{Path}\label{lowerb}

Fix a size $v\in\{z,\ldots,m\}$. We define $\lowv$ as the quantile of a set $V_v\in\Vv$, as required in \eqref{prop:Lz}, choosing $V_v$ such that it is unlikely to be rejected. We take $V_v$ as the set containing the smallest observed non-centered statistics, with the constraint that $V_v$ is an element of $\Vv$. This is a heuristic choice: $t_i$ by itself does not give full information on the rejection of $H_i$; still, if $t_i$ is small, generally $H_i$ is less likely to be rejected.

In the toy example, choose $z=1$. The set $V_v\in\mathcal{V}_1(v)$ must contain $v$ indices, at least $z=1$ of which is in $S$. Consider the observed statistics $t_i$ (first row in Table \ref{Table:toystart}, left). First, we select the column of the smallest value in $S$, then sort the remaining columns so that their values are in ascending order. Table \ref{Table:toylow} presents the centered statistics $c_i^\pi$ according to this new order. We define $V_v$ as the set of the indices of the first $v$ columns, obtaining $V_1=\{2\}$, $V_2=\{2,4\}$, $V_3=\{2,4,5\}$, $V_4=\{2,4,5,3\}$ and $V_5=M$.

\begin{table}
\caption{\label{Table:toylow} Toy example  with $S=\{1,2\}$: matrix of the sorted centered statistics to compute the path $u_1$. The value $u_1(v)$ is obtained summing the first $v$ columns by row, and then taking the quantile.}
\centering
\begin{tabular}{cccccccc}
\toprule
 &  & \multicolumn{1}{c}{selected in $S$} &  &  \multicolumn{4}{c}{remaining}\\
 &  & $i_1$ ($H_2$) &  & $j_1$ ($H_4$) & $j_2$ ($H_5$) & $j_3$ ($H_3$) & $j_4$ ($H_1$)\\
\midrule
id &  & 0 &  & 0 & 0 & 0 & 0\\
$\pi_2$ &  & 3 &  & 1 & -3 & 3 & 5\\
$\pi_3$ &  & 2 &  & -1 & 0 & 4 & -2\\
$\pi_4$ &  & 4 &  & 0 & 1 & 4 & -2\\
$\pi_5$ &  & -1 &  & 0 & -1 & 3 & 6\\
$\pi_6$ &  & 5 &  & -1 & 0 & 3 & -1\\
\bottomrule
\end{tabular}
\end{table}

In general, we select the $z$ smallest observed non-centered statistics in $S$, and then the $v-z$ remaining smallest statistics. We define two permutations of the indices:
\begin{align}
S=\{i_1,\ldots,i_s\}\quad :\quad &t_{i_1}\leq\ldots\leq t_{i_s} \label{permpath1}\\
M\setminus \{i_1,\ldots,i_z\}=\{j_1,\ldots,j_{m-z}\}\quad :\quad &t_{j_1}\leq\ldots\leq t_{j_{m-z}}.  \label{permpath2}
\end{align}
The set $\{i_1,\ldots,i_z\}$ is a subset of $S$, containing the indices of the $z$ smallest values in $S$. For instance, in the toy example we have $S=\{2,1\}$, and $M\setminus \{2\}=\{4,5,3,1\}$. The value of the path is then defined as
\begin{align}
\lowv=\qua[V_v]\qquad\text{where}\qquad V_v=\{i_1,\ldots,i_z\} \cup \{j_1,\ldots,j_{v-z}\}. \label{def:Lz}
\end{align}
It is immediate that $V_v\in\Vv$ and $\lowv\geq\upv$.

\begin{lemma}\label{L:lower}
$\min_v\lowv \leq 0$ implies $\pz=0$.
\end{lemma}

The path is used to define a function $\overline{\phi}$ as in \eqref{prop:overpz}. Similarly to the definition of $\underline{\phi}$ in the previous section, first we set $\overline{\phi}(0)=\phi(0)=0$, $\overline{\phi}(s+1)=\phi(s+1)=1$, and
\begin{align}
\overpz=\mathbf{1}\left\{\min_{v}\lowv > 0\right\}\qquad (z\in\{1,\ldots,s\}). \label{def:opz}
\end{align}
Then we make the function increasing by taking only its largest change point, imposing
\begin{align}
\overpz=0\quad\text{if}\quad \overline{\phi}(z^*)=0 \text{ for some } z^*\geq z \qquad (z\in\{1,\ldots,s\}).  \label{def:opz_bis}
%\overpz=0\quad\text{if}\quad z\leq\max\left\{\hat{z}\in\{0,\ldots,s+1\}\,:\,\overline{\phi}(\hat{z})=0 \right\}. \label{def:opz_bis}
\end{align}

\begin{prop}\label{P:short2}
$\underpz\leq\pz\leq\overpz$ for each $z\in\{0,\ldots,s+1\}$. Hence $\underpz=\overpz$ implies $\underpz=\pz$, i.e., equivalence between the shortcut and closed testing.
\end{prop}

For instance, in the toy example of Figure \ref{Plot:base2} we obtain $\underline{\phi}(1)=0<\overline{\phi}(1)=1$ and $\underline{\phi}(2)=\overline{\phi}(2)=1$. Hence the shortcut is equivalent to closed testing for $z=2$, as we already observed, but we cannot establish equivalence for $z=1$.

%To summarise, the shortcut of Proposition \ref{P:short} compares $q$ with any value $z$, then Proposition \ref{P:short2} studies whether this shortcut is equivalent to closed testing or conservative. The following section shows how to improve the shortcut in case it is conservative.

% --------------------------------------------------------------------------------------------------------------------------------------------------------------------------------------------------------------------------------------------------------------------------------------------------------------------
% --------------------------------------------------------------------------------------------------------------------------------------------------------------------------------------------------------------------------------------------------------------------------------------------------------------------

\section{Iterative shortcut}\label{bab}
The shortcut we have described in Section \ref{shortcut} approximates closed testing and efficiently computes $q^{(0)}\geq q$; however, as seen in Section \ref{equivalence}, it may be conservative. In this section we improve this single-step shortcut by embedding it into a branch and bound algorithm. We obtain an iterative shortcut which defines closer approximations of $q$, and thus smaller confidence sets for $\delta$, as the number of steps increases. Eventually, after a finite number of steps, it reaches the same results as full closed testing.

At each step $n\in\mathbb{N}$, we will define two increasing functions
\begin{align}
\underline{\phi}^{(n)},\,\overline{\phi}^{(n)}:\{0,\ldots,s+1\}\longrightarrow\{0,1\},\qquad\underpzn\leq\pz\leq\overpzn. \label{prop:pzn}
\end{align}
We will approximate $q$ from above with the change point of the first function,
\begin{align}
q^{(n)}=\max\left\{z\in\{0,\ldots,s+1\}\,:\,\underpzn =0\right\}. \label{def:qn}
\end{align}
Then we will use the second to assess possible equivalence to closed testing. If $\underpzn=\overpzn$ for a value $z$, then $\underpzn=\pz$ and so results cannot be further improved. Moreover, these functions will be defined so that $q^{(n)}$ becomes a better approximation of $q$ as $n$ increases, and finally converges to it after at most $m$ steps:
\begin{align}
& q^{(n)}\geq q^{(n+1)}\geq q^{(m)}= q \qquad (n\in\mathbb{N}).\label{prop:pzn_bis}
\end{align}

In the next sections we introduce the structure of the branch and bound algorithm, then use it to construct the functions $\underline{\phi}^{(n)}$ and $\overline{\phi}^{(n)}$ with the desired properties.

% --------------------------------------------------------------------------------------------------------------------------------------------------------------------------------------------------------------------------------------------------------------------------------------------------------------------

\subsection{Branch and bound}\label{babintr}
The branch and bound algorithm \citep{babalg, babalg2} is used when exploring a space of elements in search of a solution, and is based on the following principle. The space is partitioned into two subspaces, and each subspace is systematically evaluated; the procedure can be iterated until the best solution is found. Hence the algorithm consists of a branching rule, which defines how to generate subspaces, and a bounding rule, which gives bounds on the solution. This way, one can discard entire subspaces that, according to the bounding rule, cannot contain the solution.

Here, we want to evaluate $\pz$ for any value $z$, i.e., determine whether the space $\V$ contains a non-rejected set (see definition \eqref{def:pz}). The bounding rule that allows to make statements on the existence of such a set is the single-step shortcut of Propositions \ref{P:short} and \ref{P:short2}. If the shortcut is equivalent to closed testing, meaning that we are able to determine $\pz$, the procedure stops; otherwise, we partition $\V$ and apply the shortcut within each resulting subspace. This procedure may be iterated as needed.

\begin{figure}
\centering
\makebox{\includegraphics[width=0.9\textwidth]{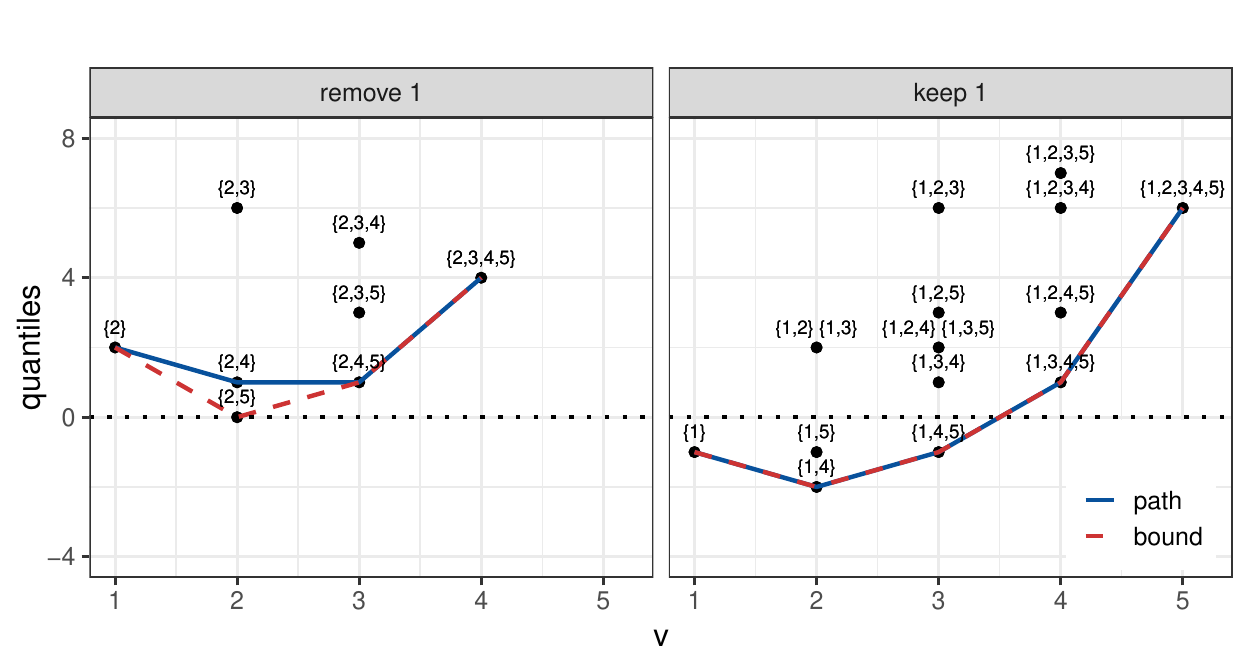}}
\caption{\label{Plot:bab} Toy example  with $S=\{1,2\}$: iterative shortcut at step $n=1$ to evaluate $\phi(z)$ in $z=1$. Points denote the quantiles for the sets in $\mathcal{V}^{-}_1$ and $\mathcal{V}^{+}_1$. The dashed and solid lines represent the bound and the path, respectively.}
\end{figure}

For instance, in the toy example, the single-step shortcut gives $\phi(2)=1$ but cannot determine $\phi(1)$ (Figure \ref{Plot:base2}). At step $n=1$, we partition $\mathcal{V}_1$ into two subspaces $\Vrem[1]$ and $\Vkeep[1]$, according to the inclusion of index $j^*=1$: $\Vrem[1]$ contains all sets that do not include $j^*$, and $\Vkeep[1]$ contains the others. We choose $j^*\in M$ as the index of the hypothesis that we believe we have most evidence against, i.e., having the greatest value $t_i$ (first row in Table \ref{Table:toystart}, left). Subsequently, we use the shortcut to examine each subspace. Figure \ref{Plot:bab} shows the bound $\up[1]$ and the path $\low[1]$ in the two subspaces; the path indicates that $\Vkeep[1]$ contains a non-rejected set, therefore we conclude that $\phi(1)=0$.

In general, the branching rule is chosen to find an eventual non-rejected set with the smallest number of steps. Fix $z\in\{1,\ldots,s\}$, as by Lemma \ref{L:phi_q} there is no need to partition $\mathcal{V}_0$ or $\mathcal{V}_{s+1}$. The space $\V$ of interest is partitioned into
\[\Vrem=\{V\in\V\,:\,j^*\notin V\},\qquad \Vkeep=\{V\in\V\,:\,j^*\in V\}\]
where $j^*$ is the index of the greatest observed non-centered statistic, with the constraint that the procedure cannot generate empty subspaces. Recall that any set $V\in\V$ has at least size $z$ overlap with $S$. Hence, with the notation of \eqref{permpath1} and \eqref{permpath2}, we fix the indices $\{i_1,\ldots,i_z\}$ of the $z$ smallest observed statistics in $S$, then we take $j^*=j_{m-z}$ as the index of the greatest remaining observed statistic. The same principle may be applied to partition any subspace.

At any step $n\in\mathbb{N}$, the procedure partitions $\V$ into $K_{n,z}$ subspaces $\V^1,\ldots,\V^{K_{n,z}}$ without any successors, where $K_{n,z}\in\{1,\ldots,2^n\}$. Suppose to apply the single-step shortcut within a subspace $\V^k$. If the result is $\pz=0$, then $\V^k$ contains a non-rejected set, and we stop with $\pz=0$. In contrast, if the shortcut determines that $\pz=1$, all sets in $\V^k$ are rejected, and we may explore other subspaces. Finally, if the shortcut produces an unsure outcome, i.e., $\pz$ is still unknown, $\V^k$ can be partitioned again.

%The bounding rule is the shortcut, as given in Propositions \ref{P:short} and \ref{P:short2}. Hence within any subspace $\V^i$ we compute $\underpz$ and $\overpz$, and use them to determine $\pz$, if possible. If $\pz=0$ in $\V^i$, then the subspace contains a non-rejected set, and the procedure stops with $\pz=0$. On the contrary, if $\pz=1$ in $\V^i$, all sets in the subspace are rejected, and we may explore further; we conclude that $\pz=1$ if this holds in all subspaces. Finally, if the shortcut produces an unsure outcome, i.e., $\pz$ is still unknown, the subspace can be partitioned again.

% --------------------------------------------------------------------------------------------------------------------------------------------------------------------------------------------------------------------------------------------------------------------------------------------------------------------

\subsection{Structure of the iterative shortcut}\label{babstr}
Fix a step $n\in\mathbb{N}$. For every $z$, the branching rule partitions $\V$ into $K_{n,z}$ subspaces $\V^1,\ldots,\V^{K_{n,z}}$, and the bounding rule applies the shortcut within them. We use this structure to define the functions $\underline{\phi}^{(n)}$ and $\overline{\phi}^{(n)}$ introduced in \eqref{prop:pzn}. We consider the point-wise minimums of $\underline{\phi}$ and $\overline{\phi}$ within the different subspaces, and so we take
\[ \underpzn = \min_k\left\{\underpz\text{ in } \V^k \right\},\qquad  \overpzn = \min_k\left\{\overpz\text{ in } \V^k \right\}.\]
Since $\underline{\phi}$ and $\overline{\phi}$ are increasing functions, also $\underline{\phi}^{(n)}$ and $\overline{\phi}^{(n)}$ are increasing. The following proposition shows that property \eqref{prop:pzn} holds, so that we can approximate $q$ from above with $q^{(n)}$, and we can assess possible equivalence to closed testing for any $z$. Moreover, the proposition gives property \eqref{prop:pzn_bis} by showing that $\underline{\phi}^{(n)}$ and $\overline{\phi}^{(n)}$ become closer to $\phi$ as $n$ increases, and finally converge to it after at most $m$ steps.

\begin{prop}\label{P:shortiter}
For any $n\in\mathbb{N}$ and any $z\in\{0,\ldots,s+1\}$,
\[\underpzn\leq\underpzn[n+1]\leq\underpzn[m]=\pz=\overpzn[m]\leq\overpzn[n+1]\leq\overpzn .\]
Hence $\underpzn=\overpzn$ implies $\underpzn=\pz$, i.e., equivalence between the iterative shortcut and closed testing. Moreover, $q^{(n)}\geq q^{(n+1)}\geq q^{(m)}=q$.
\end{prop}

%\begin{prop}\label{P:shortiter}
%$\underpzn\leq\pz\leq\overpzn$ for each $n\in\mathbb{N}$ and each $z\in\{0,\ldots,s+1\}$. Hence $q^{(n)}\geq q$. Moreover, $\underpzn=\overpzn$ implies $\underpzn=\pz$, i.e., equivalence between the iterative shortcut and closed testing.
%\end{prop}

In the toy example, consider step $n=1$  of the iterative shortcut. For $z=2$, from results of the single-step shortcut we have $\underline{\phi}^{(1)}(2)=\overline{\phi}^{(1)}(2)=\phi(2)=1$ without partitioning $\mathcal{V}_2$. For $z=1$, from Figure \ref{Plot:bab} we have $\underline{\phi}^{(1)}(1)=\overline{\phi}^{(1)}(1)=\phi(1)=0$. After one step we obtain the same results as full closed testing, with $q^{(1)}=q=1$. Then, similarly to Theorem \ref{T:d0}, at each step $n$ we may approximate $d$ from below with $d^{(n)}=s-q^{(n)}$.

\begin{theorem}\label{T:dn}
$d^{(n)}\leq d^{(n+1)}\leq d^{(m)}=d$ for each $n\in\mathbb{N}$.
\end{theorem}

Proposition \ref{P:shortiter} is the basis of the iterative shortcut. At any step $n$ and for any $z$, it allows to make statements on the value of $\pz$ by applying the single-step shortcut within at most $2^n$ subspaces. Then Theorem \ref{T:dn} gives lower ($1-\alpha$)-confidence bounds for the number of true discoveries $\delta$. Even if the iterative shortcut is stopped early, before reaching convergence, $d^{(n)}$ is always a valid lower confidence bound; we have increasingly better approximations of $d$ as $n$ increases, and obtain full closed testing results after at most $m$ steps. As the theorem may be applied to any $S \subseteq M$, the procedure $d^{(n)}$ has true discovery guarantee. In Appendix \ref{alg} we provide an algorithm for the iterative shortcut. In the worst case, the complexity of each iteration, i.e., each application of the shortcut in a subspace, is of order $mB\log(mB)$. The algorithm converges to full closed testing results after a number of iterations of order $2^m$.

% --------------------------------------------------------------------------------------------------------------------------------------------------------------------------------------------------------------------------------------------------------------------------------------------------------------------
% --------------------------------------------------------------------------------------------------------------------------------------------------------------------------------------------------------------------------------------------------------------------------------------------------------------------

\section{Refinements}\label{truncation}
In this section we show two strategies that reduce the computational time of the shortcut. First we modify the ordering of the statistics used to define the path in Section \ref{equivalence} and the branching in Section \ref{babintr}; then we introduce truncated test statistics.

Both the path and the branching are constructed sorting the indices as in \eqref{permpath1} and \eqref{permpath2}, with the intuition that a small observed value $t_i$ corresponds to a hypothesis that is less likely to be rejected. This heuristic choice may be improved if we relate the observed value with all the permuted ones, i.e., if we sort $t_i - \text{mean}(t_i^\pi)$ instead of $t_i$. This modification proved to be slightly more efficient.

%In the worst case the single-step shortcut requires a number of operations of order $m\log^2(m)$; the iterative shortcut converges to closed testing after a number of iterations exponential in $m$, where the complexity of each iteration is linearithmic in $m$.
Subsequently, recall that the computational complexity of the shortcut increases with $m$. We argue that this complexity is much reduced if the method is applied to truncated statistics, as it allows to shrink the effective total number of hypotheses from $m$ to $m'\in\{s,\ldots,m\}$. In practice, with large $B$, $m'$ is obtained by taking all statistics in $S$, and only the non-truncated observed statistics in $M\setminus S$.

Truncation-based statistics were advocated in the truncation product method of \citet{trunc0}, in the context of p-value combinations. The main idea was to emphasize smaller p-values by taking into account only p-values smaller than a certain threshold, and setting to 1 the others; a natural, common choice for the threshold is the significance level $\alpha$. A similar procedure, the rank truncation product \citep{truncrank, truncrank1}, takes into account only the $k$-th smallest p-values, for a given $k$. Eventually, weights can be incorporated into both analyses. Such procedures provide an increased power in many scenarios, and in particular for signal detection, when there is a predominance of near-null effects. They have been widely applied in literature \citep{truncapp1,truncapp2,truncapp3,truncapp4}; refer to \citet{trunc00}, \citet{trunc} and \citet{trunc2} for a review of the methods and their applications.

With our notation, we can define a truncation-based statistic for $H_S$ as following. For each hypothesis $H_i$, we set to a common ground value $\tto$ all statistics $T^\pi_i$ smaller than a threshold $\tfromi$. The threshold $\tfromi$ may depend on $i$, or be a prefixed value, or be the $k$-th greatest statistic $T_i^\pi$ ($i\in M$, $\pi\in\pib$) for a given $k$. The ground value must be $\tto\leq\min_i \tfromi$; it may be chosen, for instance, as the minimum possible value of the test statistics, or set equal to the smallest threshold $\min_i \tfromi$. Then
\[T_S = \sum_{i\in S}f_i(T_i),\qquad f_i(T_i)=\tto\cdot\mathbf{1}\{T_i<\tfromi\} + T_i\cdot\mathbf{1}\{T_i\geq\tfromi\}.\]

For simplicity of notation, let $\tfromi=\tfrom$, and so $f_i=f$, be independent of $i$. Table \ref{Table:toytrunc} shows the values $f(t_i)$ in the toy example after truncation with $\tfrom=2$ and $\tto=0$. Here, $\tfrom$ is set as the $k$-th greatest statistic, where $k=\lceil Bm\alpha\rceil$ is chosen so that the proportion of non-null contributions $f(t_i^\pi)$ is approximately $\alpha$. Observe that $H_3$ is such that the observed truncated statistic is the greatest over all permutations, i.e., $f(t_3)=\max_\pi f(t_3^\pi)$; as a consequence, adding $\{3\}$ to any set $V$ can only increase the number of rejections. On the contrary, $H_4$ and $H_5$ are such that the observed statistics are the smallest over all permutations, and so adding $\{4\}$ or $\{5\}$ to any set can only decrease rejections. Truncation makes those two particular cases more common as well as easier to check, through the following conditions:
\begin{align}
f(t_i^{\pi})&=\tto\quad\text{for all }\pi\in \pib\setminus\{\text{id}\}\label{condtrunc1}\\
f(t_i)&=\tto \label{condtrunc2}
\end{align}

\begin{table}
\caption{\label{Table:toytrunc} Toy example with $S=\{1,2\}$: test statistics after truncation of elements smaller than $\tfrom=2$ to the ground value $\tto=0$, and after dimensionality reduction.}
\centering
\begin{tabular}{ccccccccccc}
\toprule
 &  & \multicolumn{5}{c}{truncated $f(t_i^\pi)$} &  &  \multicolumn{3}{c}{dim. reduction}\\
 &  & $H_1$ & $H_2$ & $H_3$ & $H_4$ & $H_5$ &  & $H_1$ & $H_2$ & $H_{4,5}$\\
\midrule
id &  & 6 & 5 & 4 & 0 & 0 &  & 6 & 5 & 0\\
$\pi_2$ &  & 0 & 2 & 0 & 0 & 4 &  & 0 & 2 & 4\\
$\pi_3$ &  & 8 & 3 & 0 & 2 & 0 &  & 8 & 3 & 2\\
$\pi_4$ &  & 8 & 0 & 0 & 0 & 0 &  & 8 & 0 & 0\\
$\pi_5$ &  & 0 & 6 & 0 & 0 & 2 &  & 0 & 6 & 2\\
$\pi_6$ &  & 7 & 0 & 0 & 2 & 0 &  & 7 & 0 & 2\\
\bottomrule
\end{tabular}
\end{table}

\begin{prop}\label{P:trunc}
Let $V\subseteq M$ and $i\in M$. If $i$ satisfies condition \eqref{condtrunc1}, then $V\in\R$ implies $(V\cup\{i\})\in\R$. If $i$ satisfies condition \eqref{condtrunc2}, then $(V\cup\{i\})\in\R$ implies $V\in\R$.
\end{prop}

The shortcut examines the collection $\V$ of sets that have at least size $z$ overlap with $S$, searching for a set $V\notin\R$. In this case, the focus is on the number of indices in $S$, hence we may reduce the dimensionality of the problem by applying Proposition \ref{P:trunc} to the remaining indices. If an index $i\in M\setminus S$ satisfies condition \eqref{condtrunc1}, then it is not useful for finding a non-rejected set, and so can be removed from $M$. If two indices $i,j\in M\setminus S$ satisfy condition \eqref{condtrunc2}, they may be collapsed into a new index $h$, so that $H_h=H_{\{i,j\}}$ can only decrease the number of rejections. This allows to reduce the total number of hypotheses from $m$ for computational purposes to a substantially lower $m'\in\{s,\ldots,m\}$. In the toy example column 3 is removed, while columns 4 and 5 are collapsed into a single column, reducing the number of hypotheses from $m=5$ to $m'=3$.

%In Section \ref{sims} we explore simulations with different scenarios, showing the power properties of different truncation settings.

% --------------------------------------------------------------------------------------------------------------------------------------------------------------------------------------------------------------------------------------------------------------------------------------------------------------------
% --------------------------------------------------------------------------------------------------------------------------------------------------------------------------------------------------------------------------------------------------------------------------------------------------------------------

\section{Applications}\label{appl:main}
In this section, we use the iterative shortcut of Section \ref{bab} to analyse simulated and real fMRI data, while in Appendix \ref{geneappendix} we analyse differential gene expression data. We use the \texttt{sumSome} package \citep{sumSome} developed in \texttt{R} \citep{rsoftware}, with underlying code in \texttt{C++}.

% --------------------------------------------------------------------------------------------------------------------------------------------------------------------------------------------------------------------------------------------------------------------------------------------------------------------

\subsection{Simulations}\label{sims}
We use the shortcut to compare the performance of different p-value combinations through simulations. When using p-value combinations, the unknown joint distribution of the data is often managed through worst-case distributions, defined either generally or under restrictive assumptions \citep{paverage}. However, this approach makes comparisons difficult, since different tests have different worst cases. In contrast, our method adapts to the unknown distribution through permutations, and thus allows to compare the tests on equal footing. Determining which test has the highest power in different settings is a major issue, for which a full treatment is out of the scope of the paper; we present a first exploration.

We simulate $n$ independent observations from a multivariate normal distribution with $m$ variables: $\mathbf{X}=\boldsymbol{\mu}+\boldsymbol{\varepsilon}$, with $\mathbf{X},\boldsymbol{\mu},\boldsymbol{\varepsilon}\in\mathbb{R}^m$ and $\boldsymbol{\varepsilon}\sim \text{MVN}(0,\Sigma_{\rho})$. Here $\Sigma_{\rho}$ is an equicorrelation matrix with off-diagonal elements equal to $\rho$. The mean $\boldsymbol{\mu}$ has a proportion $a$ of non-null entries, with value computed so that the two-sided one-sample t-test with significance level $\alpha$ has a given power $\beta$. From the resulting data, we obtain p-values applying a two-sided one-sample t-test for each variable $i$, with null hypothesis $H_i:\,\mu_i\neq 0$. P-values are computed for $B$ random permutations. Moreover, we employ truncation, setting to a common ground value $\tto$ any p-value greater than a threshold $\tfrom$.

We analyse the subset $S$ of false hypotheses (active variables), and the complementary subset $M\setminus S$ of true hypotheses (inactive variables), by means of different p-value combinations: \citet{pearson}, \citet{liptak}, Cauchy \citep{cauchy}, and generalized means with parameter $r\in\{-2,-1,-0.5,0,1,2\}$ \citep{paverage}. The latter will be denoted by VW($r$). Notice that VW(-1) corresponds to the harmonic mean \citep{harmonic}, VW(0) to \citet{fisher}, and VW(1) to \citet{edgington}. As a comparison, we also apply the maxT-method of \citet{westyoung}, corresponding to the limit of VW($r$) when $r$ tends to $-\infty$; we apply the usual algorithm for the maxT.

We fix $n=50$, $m=1000$, $\alpha=0.05$, $B=200$ and $\tto=0.5$, then we consider $a\in\{0,0.01,0.02,0.05,0.1,0.2,0.5,0.9\}$, $\beta\in\{0.5,0.8,0.95\}$, $\rho\in\{0,0.3,0.6,0.9\}$, and $\tfrom\in\{0.005,0.01,0.05,0.1,1\}$, where $\tfrom=1$ leads to no truncation. For each setting, we simulate data 1000 times, and compute the TDP lower confidence bound for the set $S$ as the mean of $\dsa/s$ over the simulations. Furthermore, we compute the FWER as the proportion of simulations where $d(M\setminus S)>0$, meaning that the method finds at least one discovery among the true hypotheses. The algorithm is run for a maximum of 1000 iterations.

\begin{figure}
\centering
\makebox{\includegraphics[width=0.8\textwidth]{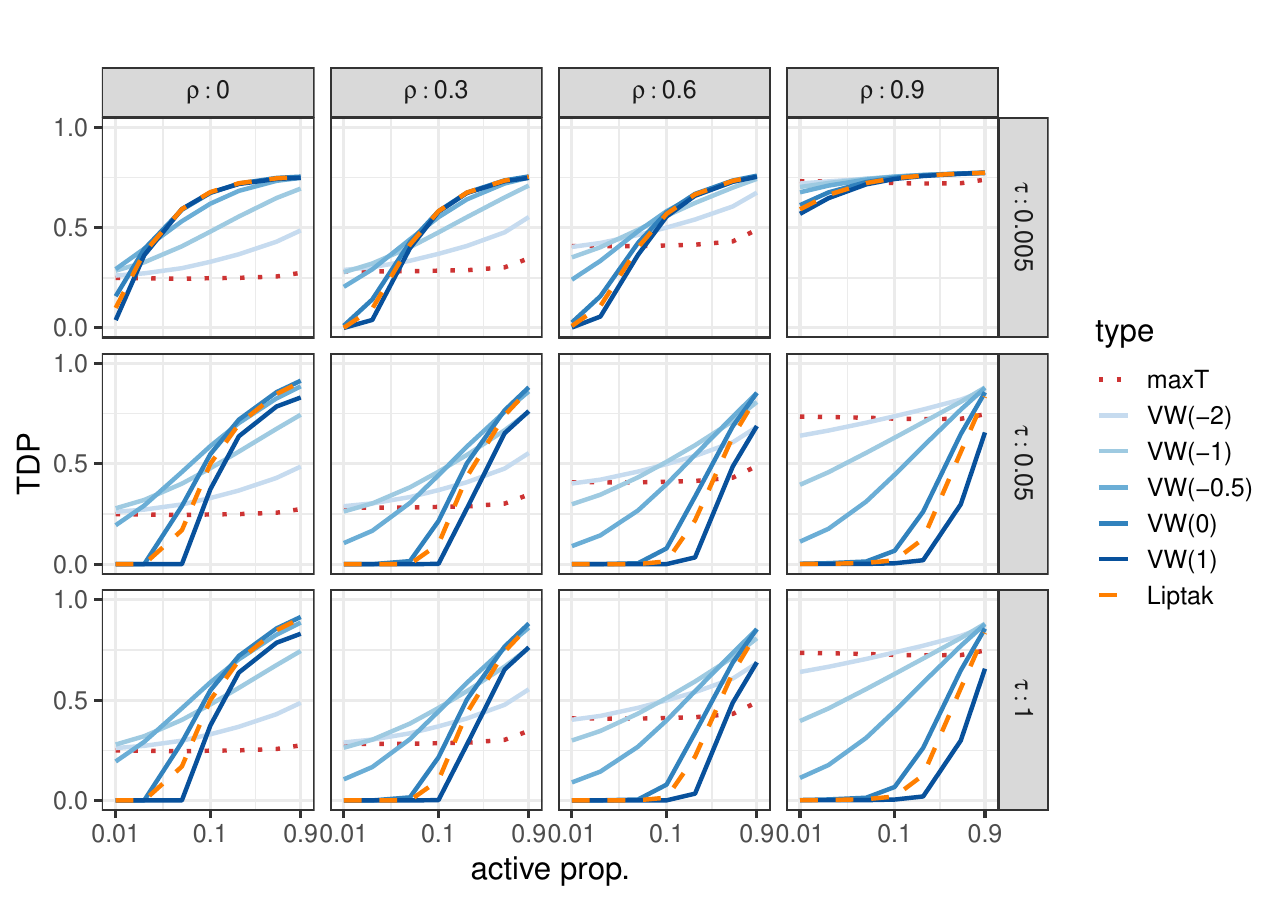}}
\caption{\label{Plot:simtdp} Simulated data: TDP lower confidence bounds for the set $S$ of active variables, by active proportion $a$ (log scale) and for different p-value combinations. Variables have equicorrelation $\rho$. P-values greater than $\tfrom$ are truncated.}
\end{figure}

Figure \ref{Plot:simtdp} shows the average TDP lower confidence bounds obtained in different scenarios for $\beta=0.95$ and $\tfrom\in\{0.005,0.05,1\}$. Certain groups of tests have similar performances: (a) VW(1), VW(2) and Pearson; (b) VW(-1) and Cauchy. For clarity, among these tests, only VW(1) and VW(-1) are displayed in the plots. Results indicate that the intensity of the signal, determined by the parameter $\beta$, does not significantly affect the behaviour of the tests; nevertheless, differences between tests are amplified when the signal is high. Furthermore, results suggest that truncation is generally advisable, unless the signal is very dense, i.e., $a$ is high. Indeed, in most cases tests tend to be more powerful when $\tfrom$ is low, and thus more statistics are truncated; the improvement is stronger for sparse signal, and when considering VW(0), VW(1) and Liptak.

When the signal is sparse, VW($r$) with $r<0$ performs best; the most powerful test is VW(-1) for low correlation, and VW(-2) for high correlation. The remaining tests perform well when the signal is dense; among those, in the considered scenarios VW(0) is the most powerful, but the powers of these tests become more similar as the signal becomes denser. These results confirm that the test is more directed towards sparse alternatives when the individual contributions, i.e., the transformed p-values, have heavy-tailed distributions, and towards dense alternatives otherwise \citep{paverage}. Computation time is between $0.04$ and 20 seconds. Moreover, simulations confirm that the method controls the FWER. Plots for the computation time, rates of convergence and the FWER are provided in Appendix \ref{fmriappendix}.

Finally, Appendix \ref{simappendix} contains a comparison with closed testing based on worst-case distributions \citep{largetdp} for generalized means VW($r$) \citep{paverage}. As expected, worst-case distributions tend to be very conservative, and are never more powerful than the shortcut. The difference in power varies according to the choice of $r$ and the setting. The largest differences are observed for $r=1$ in settings with dense signal and medium-low correlation, for which only the shortcut has non-zero power.

% --------------------------------------------------------------------------------------------------------------------------------------------------------------------------------------------------------------------------------------------------------------------------------------------------------------------

\subsection{fMRI data}\label{appl:fmri}
In this section we apply the shortcut to fMRI brain imaging data, demonstrating feasibility of the method on large datasets, adaptation to the correlation structure and post-hoc flexibility. In fMRI imaging, Blood Oxygen Level Dependent (BOLD) response is measured, i.e., changes in blood flow in the brain induced by a sequence of stimuli, at the level of small volume units called voxels. Brain activation is then inferred as correlation between the stimuli and the BOLD response. Researchers are interested in studying this activation within different clusters, brain regions of connected voxels.

Typically, voxels are highly correlated. This is usually taken into account by means of cluster extent thresholding \citep{multiplefmri, woo, ari}. However, when the method finds activation in a given cluster, it only indicates that the cluster contains at least one active voxel, but does not provide any information on the proportion of active voxels (TDP) nor their spatial location. This leads to the spatial specificity paradox, the counter-intuitive property that activation in a large cluster is a weaker finding than in a small cluster \citep{woo}. Moreover, follow-up inference inside a cluster leads to inflated Type I error rates \citep{krieg}. In contrast, our approach not only adapts to the high correlation, but also provides confidence sets for the TDP, and allows for post-hoc selection and follow-up inference inside clusters.

We analyse data collected by \citet{auditory}, which compares subjects examined while listening to vocal and non-vocal sounds. Data consists of brain images for 140 subjects, each composed of $168{,}211$ voxels. As for any standard fMRI analysis \citep{Lindquist}, as first-level analysis for each subject we estimate the contrast map that describes the difference in activation during vocal and non-vocal stimuli, with the same procedure of \citet{pARI}. Then these contrast maps are used to run the second-level analysis; for each voxel we compute a test statistic by means of a two-sided one-sample t-test, with the null hypothesis that the voxel's mean contrast between subjects is zero. Finally, we define the global test statistic for a cluster as the sum of its voxels' t-statistics.

We examine supra-threshold clusters with threshold 3.2, chosen by convention, and then we make follow-up inference inside those by studying clusters with threshold 4. The significance level is taken as $\alpha=0.05$. We construct statistics for the permutation test by using $B$ elements from the group of sign-flipping transformations, which satisfies Assumption \ref{A:perm} \citep{permglm}. Moreover, we employ truncation as in Section \ref{truncation} by setting to $\tto=0$ any statistic smaller than $\tfrom=3.2$; this way, we take into account only statistics at least as extreme as the cluster-defining threshold. We use two settings. First, we apply a ‘quick' analysis, fast and feasible on a standard machine, by using $B=200$ transformations and stopping after 50 iterations of the single-step shortcut. Subsequently, we consider a ‘long' analysis, run on the platform CAPRI \citep{capri}, that employs $B=1000$ transformations and stops after 1000 iterations. Computation time for the ‘quick' setting is less than 8 minutes on a standard PC, while the ‘long' setting requires around 9 hours for clusters with threshold 3.2, and 36 hours for follow-up inference on clusters with threshold 4.

Results, shown in Appendix \ref{fmriappendix}, indicate that the setting of the ‘long' analysis does not provide larger TDP values than the ‘quick'. Notice that the method provides valid ($1-\alpha$)-confidence bounds for the TDP in all settings. In Appendix \ref{fmriappendix} we further investigate the role of the numbers of iterations and permutations, confirming that, even though larger values give greater mean power and less variability, the ‘quick' setting provides suitable power. Moreover, our method finds activation in concordance with previous studies. An extensive comparison with other methods is beyond the scope of this paper, however our results can be immediately compared to those in \cite{pARI}, since the same data was used. For the particular settings used in the analyses, the proposed method is more powerful in detecting signal in bigger clusters, while loses power in smaller ones. In general, however, results strongly depend on the choice of the tests: the sum test in the proposed method, and the critical vector in \citet{pARI}. A preliminary study is shown in \citet{sis}.

% --------------------------------------------------------------------------------------------------------------------------------------------------------------------------------------------------------------------------------------------------------------------------------------------------------------------
% --------------------------------------------------------------------------------------------------------------------------------------------------------------------------------------------------------------------------------------------------------------------------------------------------------------------

\section{Discussion}
We have proposed a new perspective on the age-old subject of global testing, arguing that all global tests automatically come with an inbuilt selective inference method, allowing many additional inferences to be made without paying a price in terms of the global test's $\alpha$-level. Our proposed approach provides not just p-values but gives a confidence bound for the TDP, which is considerably more informative; indeed, reporting a p-value only infers the presence of some discoveries, while the TDP allows to quantify the proportion of these discoveries. Such TDP confidence bounds come not just for the full testing problem, but also simultaneously for all subsets of hypotheses; this way, subsets of interest may be chosen post hoc, without compromising the validity of the method.
%Examples of methods that make simultaneous inference on the TDP are \citet{ari} and \citet{sea}, which allow to analyse brain imaging and genomics data, respectively.

To construct simultaneous confidence bounds for the TDP of all subsets, we have provided a general closed testing procedure for sum tests, a broad class of global tests that includes many p-value combinations and other popular multiple testing methods. The procedure uses permutation testing to adapt to the unknown joint distribution of the data, avoiding strong assumptions or potential loss of power due to worst-case distributions. We have presented an iterative shortcut for this procedure, where the complexity of each iteration is linearithmic both in the numbers $m$ of hypotheses and $B$ of permutations. Moreover, we have argued that $B=200$ permutations are generally sufficient for the usual significance level $\alpha=0.05$. The shortcut converges to full closed testing results after a finite, but possibly exponential in $m$, number of iterations; furthermore, it may be stopped at any time while still providing control of the TDP. As shown in simulations, when studying 1000 hypotheses, in many cases the procedure converges to closed testing in seconds. Moreover, the method is feasible in high-dimensional settings, as shown in applications on fMRI data and differential gene expression data. An implementation is available in the \texttt{sumSome} package \citep{sumSome} in \texttt{R}, with underlying code in \texttt{C++}.

Our method is extremely flexible, allowing any sum test of choice; different choices of the sum test have very different power properties, as we have illustrated. More research is needed on the performance of different sum tests in different scenarios. Notice that the test statistic, including the eventual truncation, needs to be chosen a priori, before performing the analysis. Moreover, permutations are known to have a better performance than worst-case distributions under general dependence structure, but we have performed only a preliminary investigation to quantify the improvement given by permutations in the case of sum tests. Finally, a comparison with other permutation-based procedures that rely on bounding functions \citep{sanssouci, pARI, notip} would be of great interest, but would be extensive for two main reasons. First, all these procedures do not represent single methods but families of methods, allowing different choices for the test (i.e., sum test statistic in our case, and critical vector in the others); where and how the signal is distributed strongly influences the power of each method. Hence a fair study would require to first choose a proper test within each family, depending on many different characteristics of the problem, and only then compare results. Furthermore, the methods give statements for each of the $2^m$ possible subsets of hypotheses. Depending on the loss function chosen to summarize these statements, different methods could result to be preferable. In consequence, such an analysis is left for future work.

% --------------------------------------------------------------------------------------------------------------------------------------------------------------------------------------------------------------------------------------------------------------------------------------------------------------------
% --------------------------------------------------------------------------------------------------------------------------------------------------------------------------------------------------------------------------------------------------------------------------------------------------------------------

\bibliography{biblio}

% --------------------------------------------------------------------------------------------------------------------------------------------------------------------------------------------------------------------------------------------------------------------------------------------------------------------
% --------------------------------------------------------------------------------------------------------------------------------------------------------------------------------------------------------------------------------------------------------------------------------------------------------------------

\newpage
\appendix

\section{Algorithmic implementation}\label{alg}
In this section we provide an outline and pseudocode for the full method. In Appendix \ref{alg:main} we give the algorithms that evaluate $\pz$ for a value $z$ through the single-step and the iterative shortcut; then in Appendix \ref{bisection} we approximate $q$ by embedding these algorithms into a binary search method. Moreover, in Appendix \ref{largestset} we show how the method can be employed to find the largest subset with a given TDP. Finally, in Appendix \ref{alg:ext} we give an improved version of the algorithms of Section \ref{alg:main}.

% --------------------------------------------------------------------------------------------------------------------------------------------------------------------------------------------------------------------------------------------------------------------------------------------------------------------

\subsection{Algorithms for the shortcut}\label{alg:main}
Algorithm \ref{algorithm:short} implements the single-step shortcut of Propositions \ref{P:short} and \ref{P:short2}, evaluating $\pz$ for any given $z$ by constructing the bound $\up$ and the path $\low$. Recall that $\upv\leq\lowv$ for each size $v$. By Lemmas \ref{L:upper} and \ref{L:lower}, we have $\pz=1$ if $\up$ is entirely positive, and $\pz=0$ if $\low$ is not; in the intermediate case where $\low$ lays in the positive half-space but $\up$ does not, the value of $\pz$ remains unsure.

\begin{algorithm}
 \caption{\label{algorithm:short} Single-step shortcut to evaluate $\phi(z)$ for any $z$.}
\SetAlgoLined

\KwData{$z\in\{0,\ldots,s+1\}$}
\KwResult{$\pz$ (0, 1 or unsure)}

 \lIf{$z=0$}{\Return 0}
 \lIf{$z=s+1$}{\Return 1}
 
 Unsures = FALSE\;
 
 \For{$v=z,\ldots,m$}{
   compute $\upv$ as in (\ref{def:Uz})\;
   \If{$\upv \leq 0$}{
     Unsures = TRUE\;
     compute $\lowv$ as in (\ref{def:Lz})\;
     \lIf{$\lowv\leq 0$}{\Return 0}
    }
  }

\lIf{Unsures}{\Return unsure}
\Return 1\;
\end{algorithm}

From the result, we may obtain $\underpz$ and $\overpz$ as in (\ref{def:upz}) and (\ref{def:opz}), respectively. We do this by taking $\underpz=1$ if and only if the algorithm returns $\pz=1$, and $\overpz=0$ if and only if it returns $\pz=0$.

The procedure requires to evaluate at most $2(m-z+1)$ tests, but this number may be smaller. As shown in the following lemma, the worst-case complexity is linearithmic both in the number $m$ of hypotheses and in the number $B$ of permutations. Recall that the choice of $B$ does not depend on $m$ or $s$.

%The procedure requires to evaluate at most $2(m-z+1)$ tests, but this number may be smaller. The computational complexity is linearithmic in the number $m$ of hypotheses, and linear in the number $B$ of permutations; in the worst case, when $S=M$ and $z=1$, the total number of operations is of order $Bm\log(m)$. Recall that the choice of $B$ does not depend on $m$ or $s$.

\begin{lemma}\label{L:complexity_sumSome1}
In the worst case, the computational complexity of Algorithm \ref{algorithm:short} is of order $mB\log(mB)$.
\end{lemma}

Subsequently, Algorithm \ref{algorithm:bab} implements the iterative shortcut of Proposition \ref{P:shortiter}, embedding Algorithm \ref{algorithm:short} into a branch and bound method. First, we apply the single-step shortcut on $\V$; if it returns an unsure outcome, $\V$ is partitioned into $\Vrem$ and $\Vkeep$ as in Section \ref{babintr}. Since $\Vrem$ does not include the index $j^*$ of the hypothesis that we believe we have most evidence against, it appears more likely to contain a non-rejected set. With this reasoning, the subspaces are analysed by means of a depth-first search, meaning that the algorithm starts by exploring $\Vrem$, and explores as far as needed along the branch where indices are removed. The user can set a maximum number $h_{\max}$ of iterations, where each iteration represents the analysis of a subspace by means of the single-step shortcut. The procedure stops when it converges to closed testing results or when the number of iterations reaches $h_{\max}$.

\begin{algorithm}
 \caption{\label{algorithm:bab} Iterative shortcut to evaluate $\phi(z)$ for any $z$.}
\SetAlgoLined

\KwData{$z\in\{0,\ldots,s+1\}$; $h_{\max}\in\mathbb{N}$ (maximum number of iterations)}
\KwResult{$\pz$ (0, 1 or unsure)}

 $X=\V$\;
 shortcut on $X$ from Algorithm \ref{algorithm:short}\;
 \lIf{shortcut returns 0}{\Return 0}
 \lIf{shortcut returns 1}{\Return 1}
 $h = 0$\;
 Stack = empty list\;

\While{$h<h_{\text{max}}$}{

 \While{shortcut returns unsure \And $h<h_{\text{max}}$}{
  $++h$\;
  partition $X$ into $\Vrem$ and $\Vkeep$ as in Section \ref{babintr}\;
  add $\Vkeep$ to Stack\;
  $X=\Vrem$\;
  shortcut on $X$ from Algorithm \ref{algorithm:short}\;
  \lIf{shortcut returns 0}{\Return 0}
 }

 \While{Stack is not empty \And shortcut returns 1 \And $h<h_{\text{max}}$}{
  $++h$\;
  $X$ = last element added in Stack\;
  remove last element from Stack\;
  shortcut on $X$ from Algorithm \ref{algorithm:short}\;
 \lIf{shortcut returns 0}{\Return 0}
 }

 }

 \lIf{Stack is empty \And shortcut returns 1}{\Return 1}
 \Return unsure
\end{algorithm}

Similarly to the single-step shortcut, we obtain $\underpzn$ and $\overpzn$ by taking $\underpzn=1$ if and only if the algorithm returns $\pz=1$, and $\overpzn=0$ if and only if it returns $\pz=0$.

Each iteration of the algorithm applies the single-step shortcut of Algorithm \ref{algorithm:short} in a subspace, with complexity that is linearithmic in the number of hypotheses and the number of permutations (Lemma \ref{L:complexity_sumSome1}). The following lemma shows that the total complexity may be exponential in the number of hypotheses. In many cases this number is lower; moreover, by Theorem \ref{T:dn} we obtain a valid lower ($1-\alpha$)-confidence bound for $\delta$ even if the algorithm is stopped early.

\begin{lemma}\label{L:complexity_sumSome1bis}
In the worst case, Algorithm \ref{algorithm:bab} converges after a number of iterations of order $2^m$, where each iteration has complexity of order $mB\log(mB)$.
\end{lemma}

% --------------------------------------------------------------------------------------------------------------------------------------------------------------------------------------------------------------------------------------------------------------------------------------------------------------------

\subsection{Binary search method}\label{bisection}
Algorithms \ref{algorithm:short} and \ref{algorithm:bab} evaluate $\pz$ for a single fixed $z$, but we are interested in the change point $q$ of the function $\phi:\,\{0,\ldots,s+1\}\rightarrow\{0,1\}$, given in (\ref{def:q}). In this section, we show how to approximate $q$ without studying all values $z\in\{0,\ldots,s+1\}$.

By Lemma \ref{L:phi_q}, $\phi$ has a single change point in $q$, and takes opposite values in the extremes of its domain. Therefore $q$ may be found by embedding the shortcut within a binary search algorithm \citep{knuth}. The procedure consists of iteratively bisecting the domain of $\phi$ and selecting the subset that must contain the change point, based on the values that the function takes in the extremes and the bisection point. In the worst case, $q$ is determined after a number of steps of order $\log_2 s$. The following lemma shows that combining the single-step shortcut with a binary search has complexity at most of order $m\log^2 m$ in the number $m$ of hypotheses, and linearithmic in the number of permutations. The worst-case complexity of using the iterative shortcut of Algorithm \ref{algorithm:bab} remains exponential, as in Lemma \ref{L:complexity_sumSome1bis}.

\begin{lemma}\label{L:complexity_sumSome2}
In the worst case, embedding Algorithm \ref{algorithm:short} into a binary search requires a number of operations of order $mB(\log^2 m +\log B)$.
\end{lemma}

Even if stopped early, this procedure provides an approximation from above of $q$, and thus a valid lower ($1-\alpha$)-confidence bound for the number of true discoveries $\delta$. For instance, if we stop after finding that $q\in\{z_1,\ldots,z_2\}$, we know that $q \leq z_2$, and so $d\geq s-z_2$. As a result, $s-z_2$ is a lower ($1-\alpha$)-confidence bound for $\delta$.

% --------------------------------------------------------------------------------------------------------------------------------------------------------------------------------------------------------------------------------------------------------------------------------------------------------------------

\subsection{Largest subset with given TDP}\label{largestset}
In this section we show how the shortcut can be used to study incremental sets in any desired ordering, and quickly determine the largest set having TDP at least equal to a given value $\eta$. This is achieved by combining the binary search of Section \ref{bisection} with an algorithm of \citet{largetdp}.

In the toy example, suppose we want to study the incremental sets $S_1=\{1\}$, $S_2=\{1,2\}$,  $S_3=\{1,2,3\}$,  $S_4=\{1,2,3,4\}$ and $S_5=M$, with the aim of finding the largest one having TDP lower confidence bound at least equal to $0.5$. This means finding the greatest size $s$ such that $d(S_s)/s\geq 0.5$; the values $d(S_s)$ can be computed by means of the binary search.

In general, let $S_1\subset\ldots\subset S_m=M$ be a collection of incremental sets, with size $|S_s|=s$ for each $s$, and fix $\eta\in [0,1]$. Then Algorithm \ref{algorithm:largest} finds the greatest size $s$ with $d(S_s)/s \geq\eta$. From the result, we know that the TDP of the selected set is at least $\eta$ with confidence $1-\alpha$.

\begin{algorithm}
\caption{\label{algorithm:largest} Procedure to study a collection of incremental sets, and determine the size of the largest set with TDP lower confidence bound at least equal to $\eta$.}
\SetAlgoLined

\KwData{$S_1\subseteq\ldots\subseteq S_m$ (incremental sets with $|S_s|=s$ for each $s$); $\eta\in [0,1]$}
\KwResult{$\max\left\{s\in\{1,\ldots,m\}\,:\,d(S_s)/s \geq\eta\right\}$ if the maximum exists, 0 otherwise}

 \lIf{$\eta = 0$}{\Return m}

 s = m\;
 \While{$s>0$}{
  binary search on $S_s$ to compute $d(S_s)$\;
  \lIf{$d(S_s)/s \geq\eta$}{\Return s}
  $s=\lfloor d(S_s)/\eta\rfloor$\;
 }

 \Return 0
\end{algorithm}

% --------------------------------------------------------------------------------------------------------------------------------------------------------------------------------------------------------------------------------------------------------------------------------------------------------------------

\subsection{Algorithms for the shortcut with reduced complexity}\label{alg:ext}
In this section, we provide improved versions of Algorithms \ref{algorithm:short} and \ref{algorithm:bab}, which require fewer computations. Algorithms \ref{algorithm:short2} and \ref{algorithm:bab2} evaluate $\pz$ for any $z$, using the single-step shortcut of Propositions \ref{P:short} and \ref{P:short2} and the iterative shortcut of Proposition \ref{P:shortiter}, respectively. The number of computations is reduced by exploiting some properties of the bound and the path.

\begin{algorithm}
 \caption{\label{algorithm:short2} Single-step shortcut to evaluate $\phi(z)$ for any $z$ (algorithm with reduced complexity).}
\SetAlgoLined
\KwData{$z\in\{0,\ldots,s+1\}$; $v_1,v_2\in\{z,\ldots,m\}$ (smallest and greatest sizes to check); get\_path (TRUE to compute the path)}
\KwResult{$\pz$ (0, 1 or unsure); if $\pz$ is unsure, $v_1$ and $v_2$ are updated}

 \lIf{$z=0$}{\Return 0}
 \lIf{$z=s+1$}{\Return 1}
 
 compute $\ci[1]$ and $\ci[2]$ as in (\ref{def:c1}) and (\ref{def:c2})\;
 Unsures = empty list;
 
 \For{$v=\ci[1], \ci[1] -1\ldots,v_1$}{
   compute $\upv$ as in (\ref{def:Uz})\;
   \lIf{$\upv > 0$}{\Break}
   add $v$ to Unsures\;
   \If{get\_path}{
      compute $\lowv$ as in (\ref{def:Lz})\;
      \lIf{$\lowv\leq 0$}{\Return 0}
    }
 }
  
  \For{$v=v_1+1, v_1+2,\ldots,v_2$}{
   compute $\upv$ as in (\ref{def:Uz})\;
   \lIf{$\upv > 0$ \And $v\geq \ci[2]$}{\Break}
   \uElseIf{$\upv \leq 0$}{
    add $v$ to Unsures\;
    \If{get\_path}{
      compute $\lowv$ as in (\ref{def:Lz})\;
      \lIf{$\lowv\leq 0$}{\Return 0}
    }
   }
  }
 
\lIf{Unsures is empty}{\Return 1}
update $v_1=\min\text{Unsures}$ and $v_2=\max\text{Unsures}$\;
\Return unsure\;
\end{algorithm}

\begin{algorithm}
 \caption{\label{algorithm:bab2} Iterative shortcut to evaluate $\phi(z)$ for any $z$ (algorithm with reduced complexity).}
\SetAlgoLined

\KwData{$z\in\{0,\ldots,s+1\}$; $h_{\max}$ (maximum number of iterations)}
\KwResult{$\pz$ (0, 1 or unsure)}

 $X=\V$\;
 $v_1=z$; $v_2=m$\;
 get\_path = TRUE\;
 shortcut on $X$ from Algorithm \ref{algorithm:short2}\;
 \lIf{shortcut returns 0}{\Return 0}
 \lIf{shortcut returns 1}{\Return 1}
 $h = 0$\;
 Stack = empty list\;
\While{$h<h_{\text{max}}$}{

  get\_path = FALSE\;

 \While{shortcut returns unsure \And $h<h_{\text{max}}$}{
  $++h$\;
  partition $X$ into $\Vrem$ and $\Vkeep$ as in Section \ref{babintr}\;
  add ($\Vkeep$, $v_1$, $v_2$) to Stack\;
  $X=\Vrem$\;
  shortcut on $X$ from Algorithm \ref{algorithm:short2}\;
  \lIf{shortcut returns 0}{\Return 0}
 }
 
 get\_path = TRUE\;

 \While{Stack is not empty \And shortcut returns 1 \And $h<h_{\text{max}}$}{
  $++h$\;
  ($X$, $v_1$, $v_2$) = last element added in Stack\;
  remove last element from Stack\;
  shortcut on $X$ from Algorithm \ref{algorithm:short2}\;
 \lIf{shortcut returns 0}{\Return 0}
 }

 }

 \lIf{Stack is empty \And shortcut returns 1}{\Return 1}
 \Return unsure
\end{algorithm}

The structure of Algorithm \ref{algorithm:short2} is constructed so that it will be useful for Algorithm \ref{algorithm:bab2}. The user can define for which sizes $v\in\{v_1,\ldots,v_2\}\subseteq\{z,\ldots,m\}$ they need to check whether $\Vv\subseteq\R$ (see partition (\ref{def:partition})), supposing this is already known to be true for the remaining sizes. The algorithm not only evaluates $\pz$, but also updates the values of $v_1$ and $v_2$, keeping track of the new set, possibly empty, of sizes that need to be further examined. Moreover, the user can state whether they already know that the path $\low$ is entirely positive; in this case, it is not computed.

We reduce the number of computations needed by Algorithm \ref{algorithm:short2} as follows. In Lemma \ref{L:upper_shape} we will show that there exist $\ci[1],\ci[2]\in\{z,\ldots,m\}$ such that the bound $\upv$ is decreasing for $v\leq \ci[1]$, and increasing for $v\geq \ci[2]$. Since the single-step shortcut only uses the sign of $\min_v\upv$ (Lemma \ref{L:upper}), it is not always necessary to compute $\upv$ for all sizes $v$. For instance, in the toy example with $z=1$ (Figure \ref{Plot:base2}, left), we have $\ci[1]=\ci[2]=2$; since $\up[1](4)>0$ and $4\geq \ci[2]$, we know that $\up[1](5)>0$ without computing it.

We find $\ci[1]$ and $\ci[2]$ as following. Consider the toy example with $z=1$, and recall definition (\ref{def:Uz}). The value $\up[1](v)$ is computed by summing by row the first $v$ columns of Table \ref{Table:toyup}, and then taking the quantile. Note that the elements of column 3 are all non-negative; as a consequence, $b_2^\pi\leq b_3^\pi$ for each $\pi$, and so $\up[1](2)\leq \up[1](3)$. Moreover, since the statistics are sorted so that $c_{j_1(\pi)}^\pi\leq \ldots\leq c_{j_4(\pi)}^\pi$ for each $\pi$, all elements of columns 4 and 5 must be non-negative, and so $\up[1](2)\leq\ldots\leq\up[1](5)$. A similar argument may be used to show that, since all elements of column 2 are non-positive, $\up[1](1)\geq\up[1](2)$.

In general, fix $z\in\{1,\ldots,s\}$, and consider the statistics $c_{j_h(\pi)}^\pi$, with $h\in\{1,\ldots,m-z\}$, that remain after selecting the $z$ smallest statistics in $S$ for each $\pi$. We identify the last column where these statistics are all non-positive, taking
\begin{align}
\ci[1] = z + \max\left\{h\in\{1,\ldots,m-z\}\,:\,c_{j_h(\pi)}^{\pi}\leq 0\;\text{for all }\pi\in\pib\right\} \label{def:c1}
\end{align}
if the maximum exists, and $\ci[1]=z$ otherwise. Similarly, we identify the last column before these statistics become all non-negative, taking
\begin{align}
\ci[2] = z + \max\left\{h\in\{1,\ldots,m-z\}\,:\,c_{j_h(\pi)}^{\pi} < 0\;\text{for some }\pi\in\pib\right\} \label{def:c2}
\end{align}
if the maximum exists, and $\ci[2]=z$ otherwise.

\begin{lemma}\label{L:upper_shape}
Define $\ci[1]$ as in (\ref{def:c1}), and $\ci[2]$ as in (\ref{def:c2}). Then $\up(z)\geq\up(z+1)\geq\ldots\geq\up(\ci[1])$, and $\up(\ci[2])\leq \up(\ci[2] + 1)\ldots\leq\up(m)$.
\end{lemma}

Subsequently, Algorithm \ref{algorithm:bab2} embeds Algorithm \ref{algorithm:short2} within a branch and bound method. With respect to Algorithm \ref{algorithm:bab}, the number of computations is reduced as following. First, Lemma \ref{L:b_in_sub} shows that it is not necessary to compute the path $\low$ when we apply the single-step shortcut within $\Vrem$; where it is defined, it coincides with the path in $\V$ (e.g., compare Figures \ref{Plot:base2} and \ref{Plot:bab}). The same argument applies any time an index is removed.

\begin{lemma}\label{L:b_in_sub}
$\lowv$ is the same in $\V$ and $\Vrem$ for each $v\in\{z,\ldots,m-1\}$.
\end{lemma}

Moreover, when studying subspaces it is only necessary to further examine those sizes $v$ for which we are unsure whether $\Vv\subseteq\R$. For instance, in the toy example with $z=1$, the single-step shortcut gives $\up[1](4),\, \up[1](5)>0$ (Figure \ref{Plot:base2}, left), and so we only need to further examine $v\in\{1,2,3\}$ when applying the single-step shortcut within $\Vrem[1]$ and $\Vkeep[1]$.

% --------------------------------------------------------------------------------------------------------------------------------------------------------------------------------------------------------------------------------------------------------------------------------------------------------------------
% --------------------------------------------------------------------------------------------------------------------------------------------------------------------------------------------------------------------------------------------------------------------------------------------------------------------

\section{Applications}\label{applappendix}
In this section, we provide additional information on the applications of Section \ref{appl:main}, as well as another application on real gene expression data. First, we give results on the analysis of simulated data. Then we show results for the application on fMRI data, investigating the impact of the number of iterations and permutations on power. Finally, we analyse differential gene expression data.

% --------------------------------------------------------------------------------------------------------------------------------------------------------------------------------------------------------------------------------------------------------------------------------------------------------------------

\subsection{Simulations}\label{simappendix}
We give results for the simulations of Section \ref{sims}. First, we show the computation time, convergence rates and the FWER of the shortcut in scenarios with $\beta=0.95$ and $\tfrom\in\{0.005,0.05,1\}$; other values of $\beta$ and $\tfrom$ lead to analogous results. Subsequently, we compare the shortcut with closed testing based on worst-cased distributions for generalized means VW($r$) \citep{paverage}, using an algorithm of \citet{largetdp}.

\begin{figure}
\centering
\makebox{\includegraphics[width=0.8\textwidth]{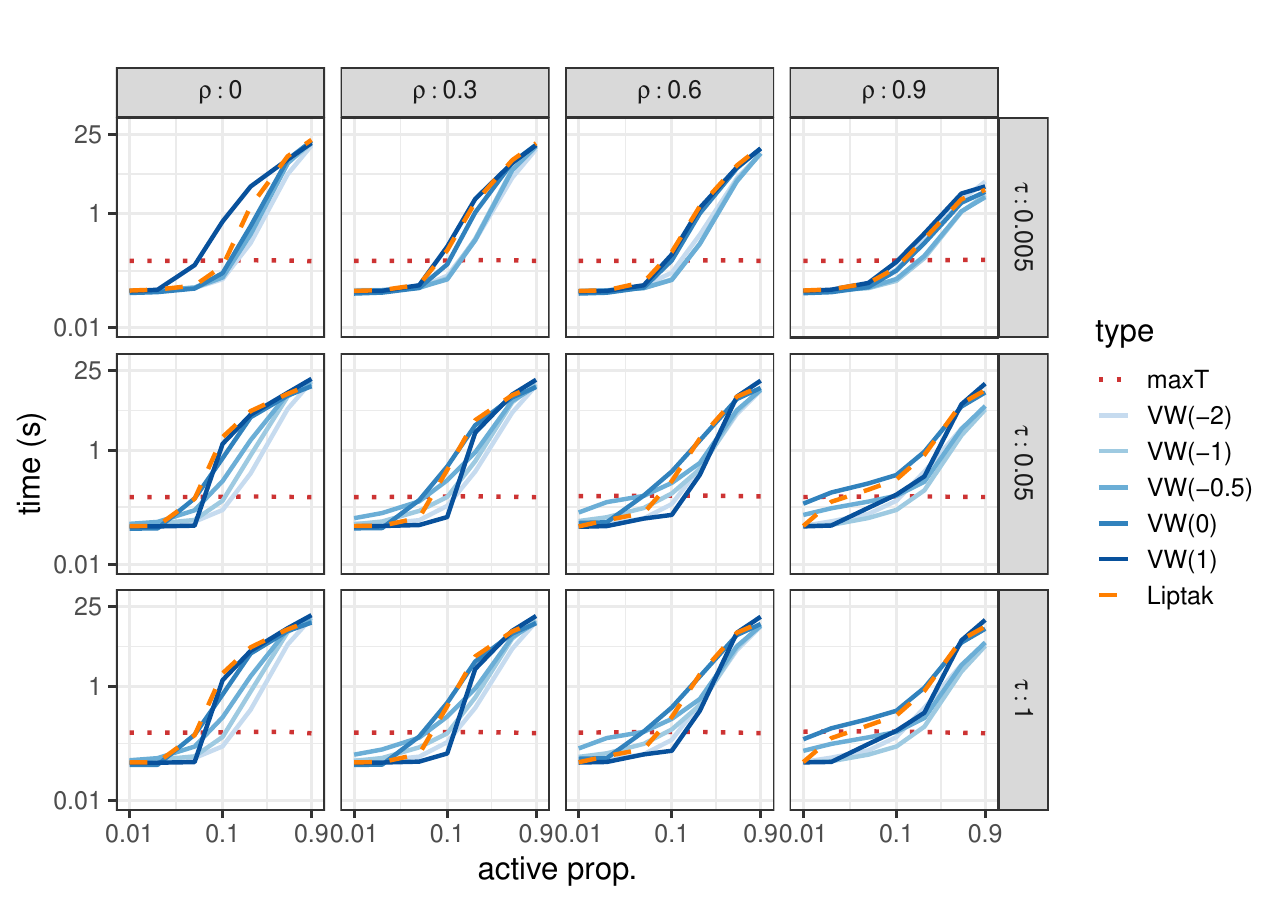}}
\caption{\label{Plot:simtime} Simulated data: computation time (log scale) for the analysis of the set $S$ of active variables, by active proportion $a$ (log scale) and for different p-value combinations. Variables have equicorrelation $\rho$. P-values greater than $\tfrom$ are truncated.}
\end{figure}

\begin{figure}
\centering
\makebox{\includegraphics[width=0.8\textwidth]{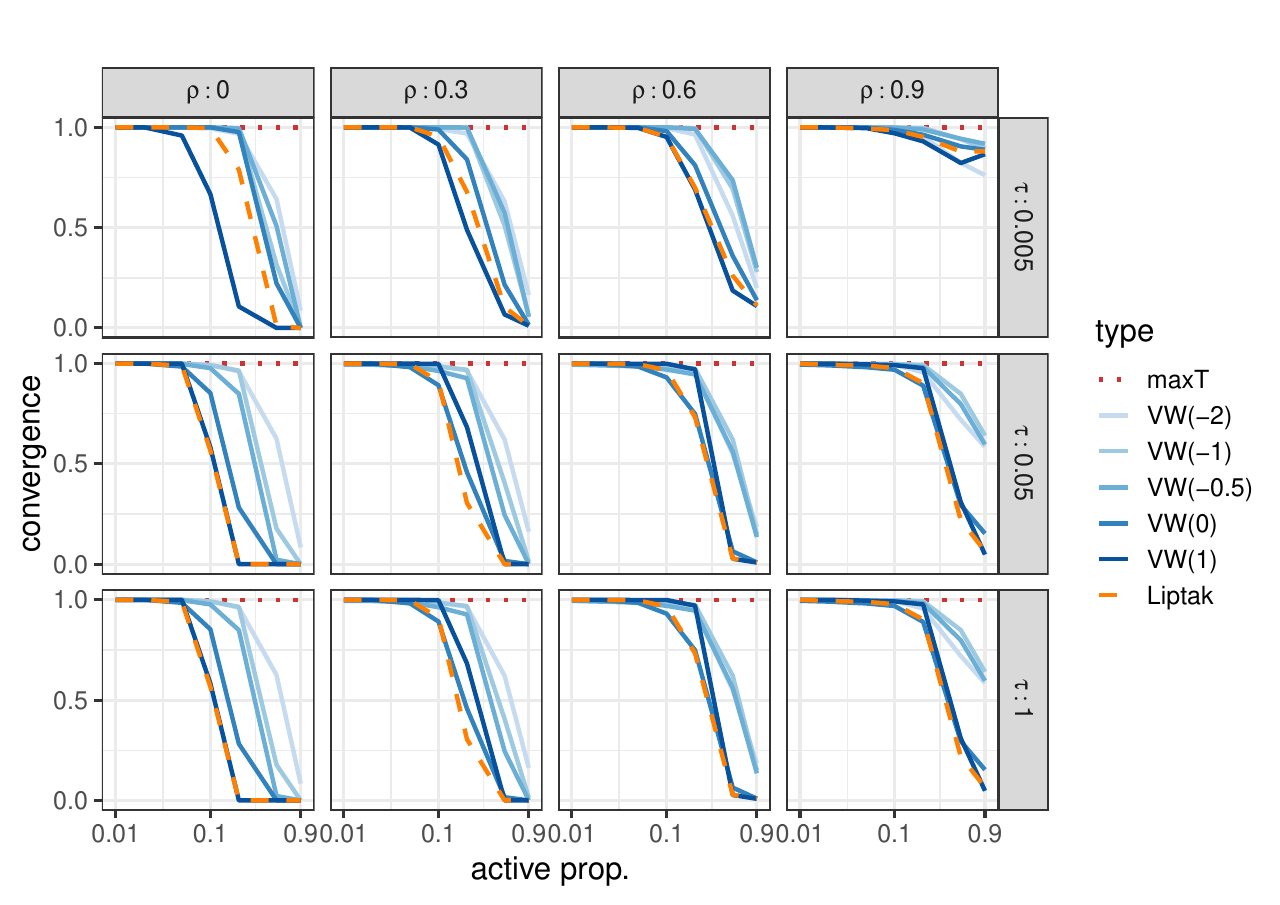}}
\caption{\label{Plot:simconv} Simulated data: convergence to full closed testing for the analysis of the set $S$ of active variables, by active proportion $a$ (log scale) and for different p-value combinations. Variables have equicorrelation $\rho$. P-values greater than $\tfrom$ are truncated.}
\end{figure}

Figures \ref{Plot:simtime} and \ref{Plot:simconv} display the computation time needed to analyse the set $S$ of false hypotheses and the convergence rates of the algorithm to full closed testing results. Time is always less than $20$ seconds, and increases with the size of the considered set $S$, i.e., with $a$. Convergence rates are low when the set $S$ is big and less p-values are truncated, i.e., when $a$ and $\tfrom$ are high.

\begin{figure}
\centering
\makebox{\includegraphics[width=0.8\textwidth]{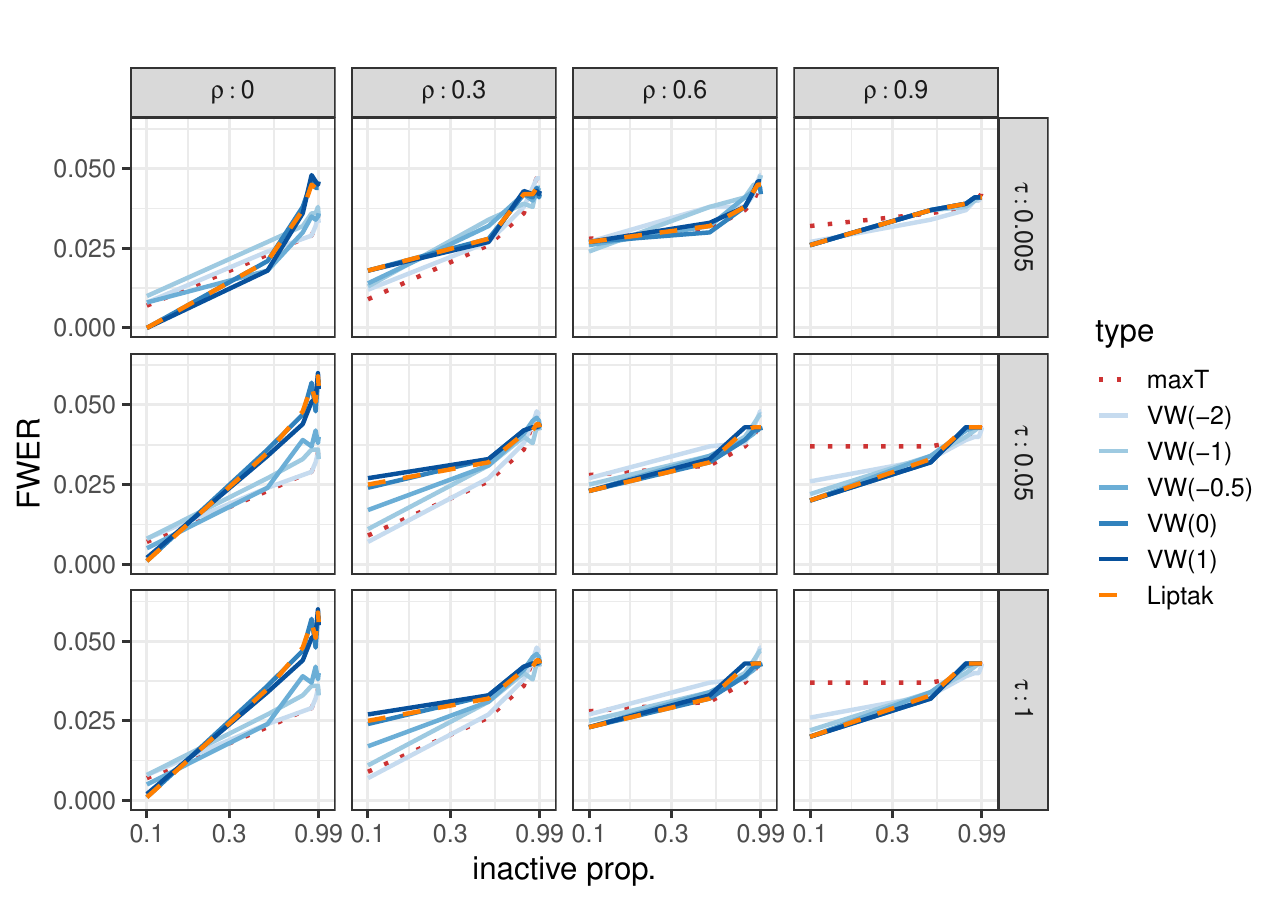}}
\caption{\label{Plot:simfwer} Simulated data: FWER computed on the set $M\setminus S$ of inactive variables, by inactive proportion $1-a$ (log scale) and for different p-value combinations. Variables have equicorrelation $\rho$. P-values greater than $\tfrom$ are truncated.}
\end{figure}

Figure \ref{Plot:simfwer} shows the FWER, computed as the proportion of times when the method finds at least one discovery among the set $M\setminus S$ of true hypotheses. Results confirm that the procedure controls the FWER; indeed, the FWER never exceeds the significance level $\alpha$ by more than two standard deviations, i.e., it is never higher than $0.063$.

\begin{figure}
\centering
\makebox{\includegraphics[width=0.8\textwidth]{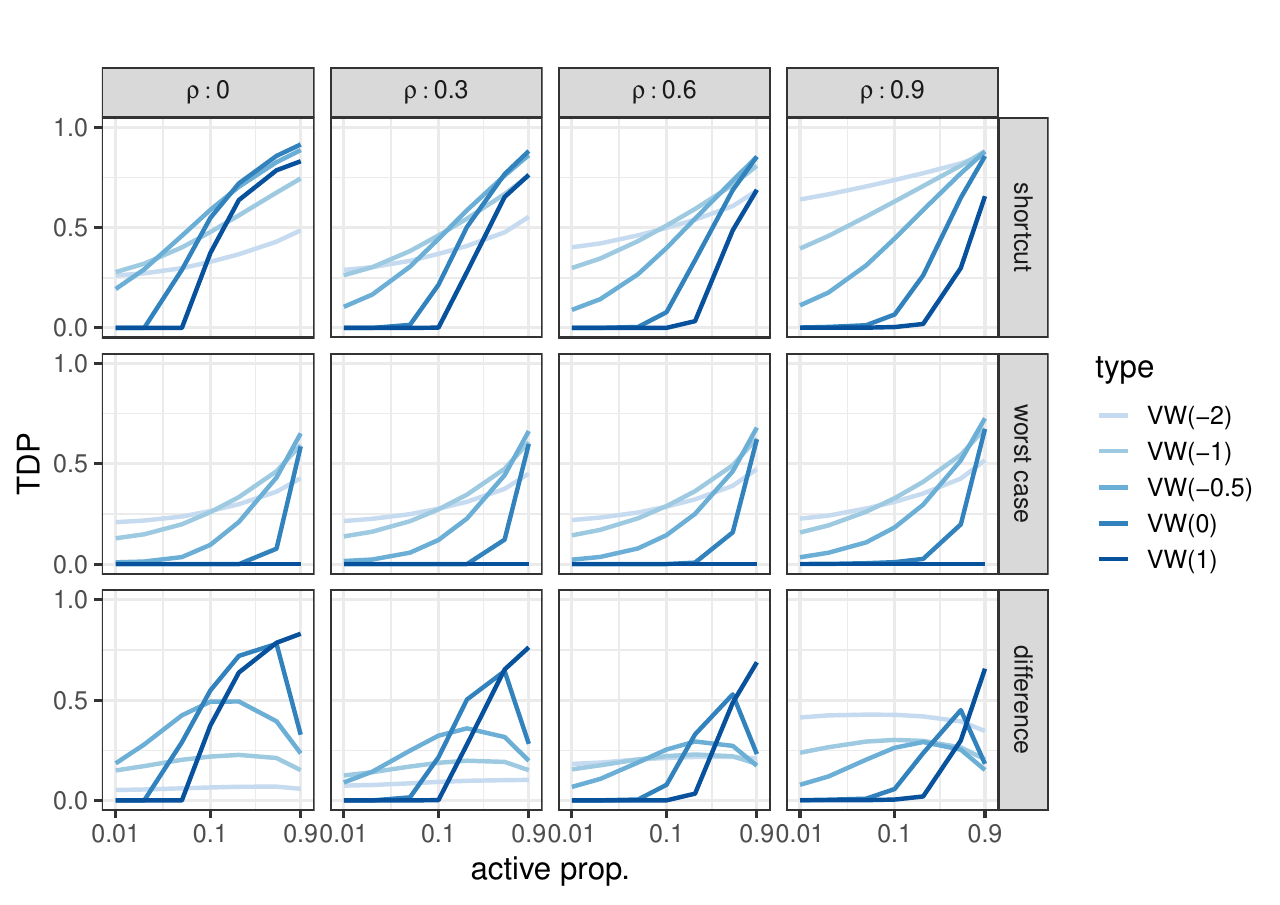}}
\caption{\label{Plot:simcomp} Simulated data: TDP lower confidence bounds for the set $S$ of active variables, by active proportion $a$ (log scale) and for different p-value combinations. Variables have equicorrelation $\rho$. Results are shown for the proposed shortcut, for closed testing based on worst-case distributions, and as a difference between the two methods.}
\end{figure}

Finally, we consider the comparison between the shortcut and closed testing based on worst-case distributions, focusing on generalized means  VW($r$) \citep{paverage} with no truncation ($\tfrom=1$). Figure \ref{Plot:simcomp} shows the TDP obtained from both methods, as well as the difference between the TDP given by the shortcut and that given by worst-case distributions. Interestingly, the two methods exhibit a similar behaviour, but the shortcut is always at least as powerful as worst-case distributions. The difference in performance varies with the choice of the test as well as with the setting.

% --------------------------------------------------------------------------------------------------------------------------------------------------------------------------------------------------------------------------------------------------------------------------------------------------------------------

\subsection{fMRI data}\label{fmriappendix}
\begin{figure}
\centering
\makebox{\includegraphics[width=0.9\textwidth]{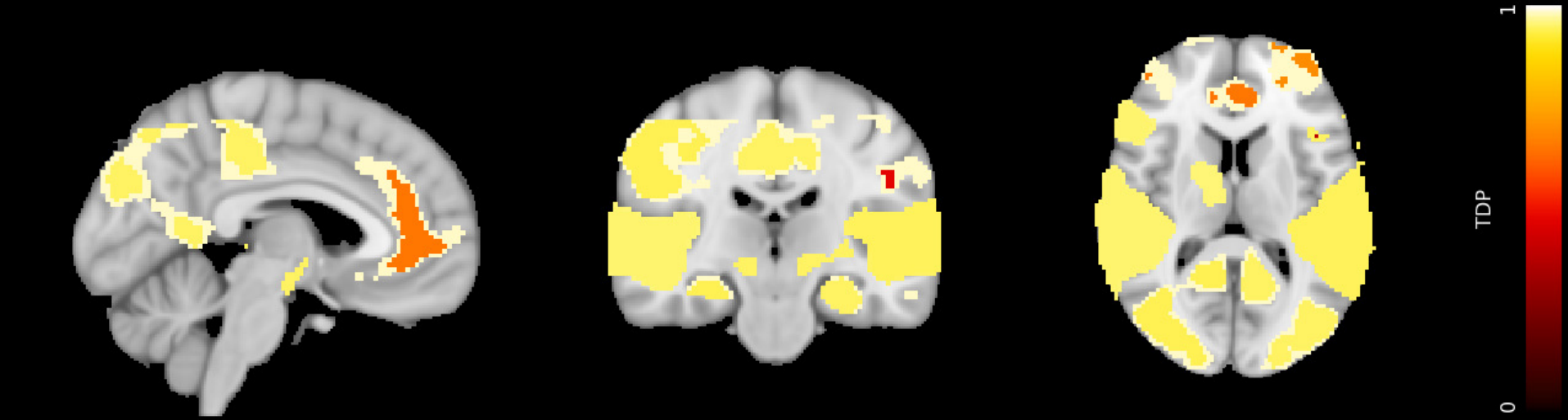}}
\caption{\label{Plot:fmri2} fMRI data: map of the TDP lower confidence bounds for supra-threshold clusters with thresholds 3.2 and 4.}
\end{figure}

Results for the analysis of Section \ref{appl:fmri} are shown in Table \ref{Table:fmri2}, which contains the lower confidence bound for the TDP of each cluster, as well as the size and the coordinates of the maximum t-statistic. Figure \ref{Plot:fmri2} contains the map of the TDP lower confidence bounds obtained from the ‘quick' setting. Results indicate that the setting of the ‘long' analysis does not provide larger TDP values than the ‘quick'. While in this particular case `quick' calculation settings tend to give slightly better results, the difference is dominated by the variability due to the random permutations.

\begin{table}
\caption{\label{Table:fmri2} fMRI data: analysis of supra-threshold clusters with thresholds 3.2 and 4. Clusters with no discoveries are not shown.}
\centering
%\begin{adjustbox}{max width=\textwidth}
\begin{tabular}{rccccccc}
\toprule
\multicolumn{1}{c}{cluster} & threshold & size & \multicolumn{2}{c}{TDP ($\%$)} & \multicolumn{3}{c}{coordinates}\\
\multicolumn{1}{c}{$S$} & $thr$ & $s$ & \multicolumn{2}{c}{$\dsa/s$ ($\%$)} & $x$ & $y$ & $z$ \\ 
 &  &  & quick & long &  &  &  \\ 
\midrule
\rowcolor{lightgray} FP/CG/SFG/TOF/LO/LG/ & 3.2 & $40{,}094$ & $98.23$ & $97.81$ & -30 & -34 &	-16 \\
\rowcolor{lightgray} OFG/ITG/SG/AG/NA &  &  &  &  &  &  &  \\
\rowcolor{white} Left LO/TOF & 4 & $8{,}983$ & $95.00$ & $93.86$ & -30 & -34 & -16 \\
\rowcolor{white} Right LO/LG/ITG & 4 & $7{,}653$ & $94.07$ & $92.94$ & 28 & -30 & -18 \\
\rowcolor{white} Left SFG/FP & 4 & $1{,}523$ & $69.93$ & $66.91$ & -28 & 34 & 42 \\
\rowcolor{white} CG & 4 & $1{,}341$ & $67.41$ & $62.94$ & 6 & 40 & -2 \\
\rowcolor{white} Right FP & 4 &  $1{,}327$  & $66.16$ & $62.40$ & 30 &	56 &	28 \\
\rowcolor{white} Left SG/AG & 4 & $859$ & $49.83$ & $43.66$ & -50 & -56 & 36 \\
\rowcolor{lightgray} Right STG/PT/MTG/ & 3.2 & $12{,}540$ & $95.35$ & $95.02$ & 60 & -10 & 0 \\
\rowcolor{lightgray} HG/PrG/T &  &  &  &  &  &  &  \\
\rowcolor{white} STG/PT/MTG/HG & 4 & $9{,}533$ & $95.30$ & $94.79$ & 60 & -10 & 0 \\
\rowcolor{white} PrG & 4 & $485$ & $27.42$ & $20.41$ & 52 & 0 &	48 \\
\rowcolor{lightgray} Left STG/PT/MTG/& 3.2 & $10{,}833$ & $94.70$ & $94.08$ & -60 & -12 & 2 \\
\rowcolor{lightgray} HG/IFG/T &  &  &  &  &  &  &  \\
\rowcolor{white} HG/PT/MTG/STG & 4 & $7{,}894$ & $94.34$ & $93.55$ & -60 &	-12 & 2 \\
\rowcolor{white} IFG & 4 & $667$ & $41.23$ & $34.78$ & -40 & 14 & 26 \\
\bottomrule
\end{tabular}
%\end{adjustbox}
\end{table}

Subsequently, we investigate the role of the number of iterations of the single-step shortcut and the number of permutations in the analysis. We do this by examining two clusters: $(1)$ the biggest cluster with threshold 3.2 (FP/CG/SFG/TOF/LO/LG); $(2)$ its smallest sub-cluster with non-null activation (Left SG/AG).

\begin{figure}
\centering
\makebox{\includegraphics[width=0.9\textwidth]{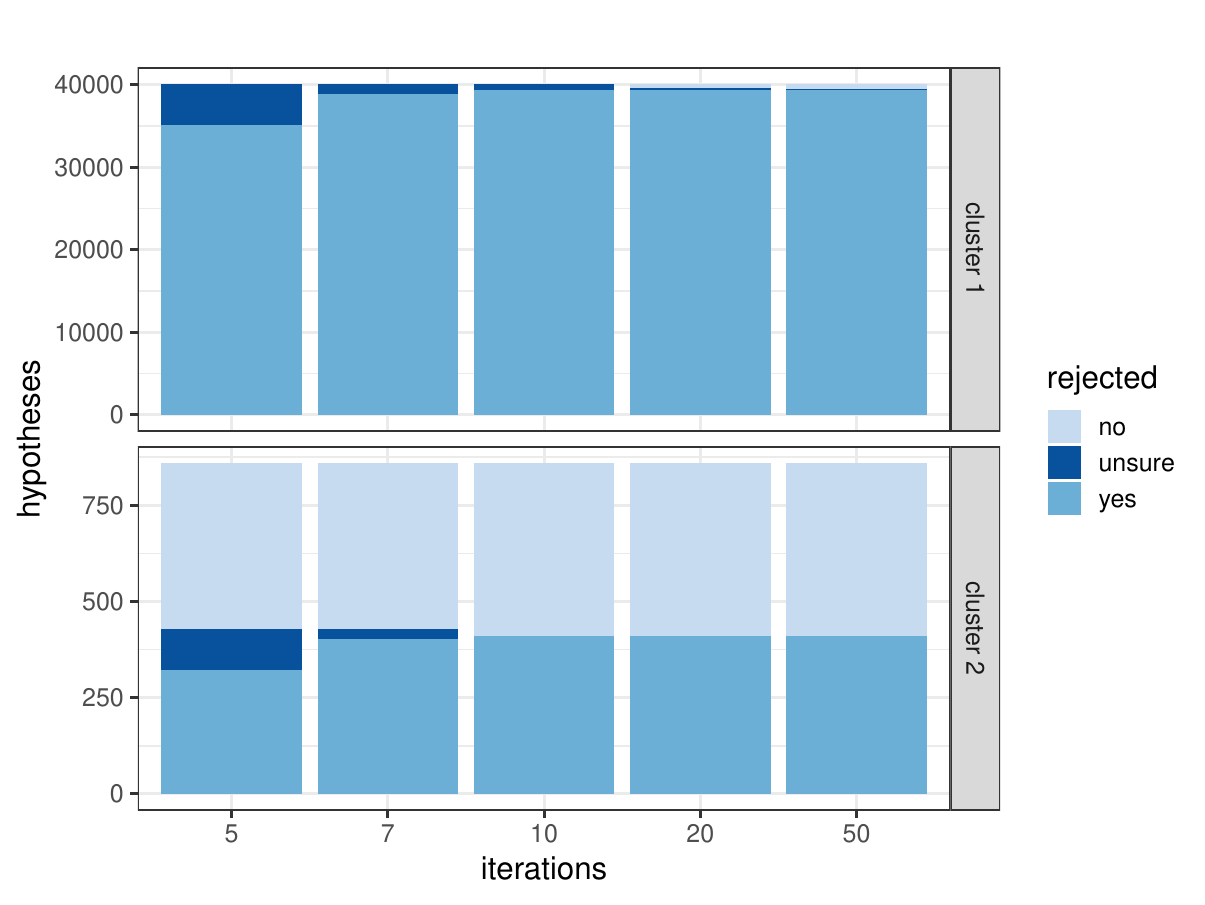}}
\caption{\label{Plot:tdptime} fMRI data: rejected, non-rejected and unsure hypotheses by number of iterations, for clusters (1) FP/CG/SFG/TOF/LO/LG; (2) Left SG/AG.}
\end{figure}

First, we analyse these two clusters, stopping the algorithm at different times; the number of permutations is fixed at $B=200$. Figure \ref{Plot:tdptime} shows the number of rejected, non-rejected and unsure hypotheses by the number of iterations. As expected, the number of unsure hypotheses quickly decreases, and becomes less than $0.5\%$ of the total after only 20 iterations. A high number of iterations is required only for the very last unsure hypotheses.

\begin{figure}
\centering
\makebox{\includegraphics[width=0.9\textwidth]{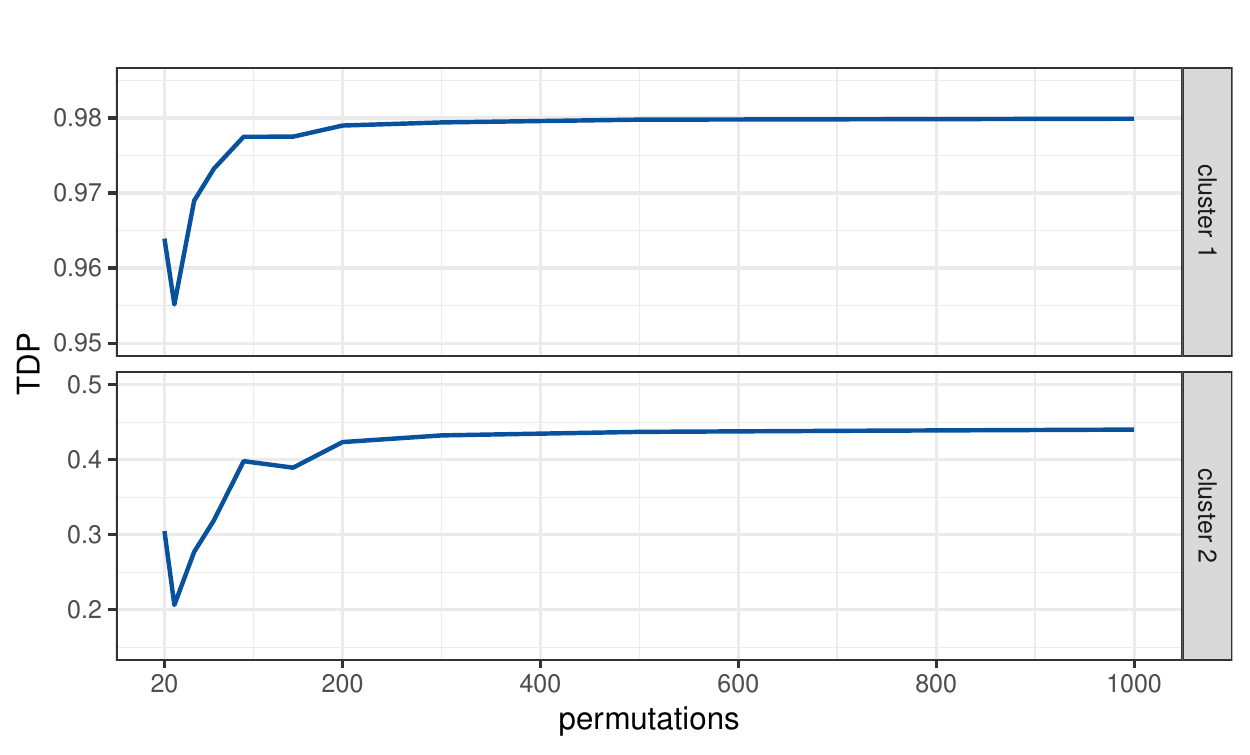}}
\caption{\label{Plot:tdpb} fMRI data: TDP lower confidence bounds by number of permutations, for clusters (1) FP/CG/SFG/TOF/LO/LG; (2) Left SG/AG.}
\end{figure}

Subsequently, to investigate the impact of the number $B$ of permutations on power, we study the clusters with different values $B$. Figure \ref{Plot:tdpb} shows the mean TDP lower confidence bound by $B$, obtained by performing each analysis 1000 times and using at most 50 iterations. The power is increasing for $B\leq 300$, and then becomes approximately constant; from $B=200$ to $300$, the gain is very small. Notice that the power peaks when $B$ is a multiple of $1/\alpha$, since the permutation test is exact only for these values of $B$ \citep{exact}. Aside from a lower mean power, another drawback of using few permutations is that results may be variable, due to the randomness of the permutations.

% --------------------------------------------------------------------------------------------------------------------------------------------------------------------------------------------------------------------------------------------------------------------------------------------------------------------

\subsection{Differential gene expression data}\label{geneappendix}
In this section we analyse differential gene expression data, obtained quantifying the differences in the levels of each gene's product (generally a protein) between two populations. Researchers often want to assess differences at the level of pathways, i.e., collections of genes associated with a specific biological process that interact with each other. The features of this analysis are analogous to those illustrated in Section \ref{appl:fmri}: genes are correlated with each other, and interest lies in studying multiple gene sets.

We consider the Breast Invasive Carcinoma dataset from \citet{brca}, containing gene expression data for patients with different types of breast tumor. We select patients with primary solid tumor and study differences between populations defined by the following histological types: infiltrating lobular carcinoma and infiltrating ductal carcinoma. Moreover, we filter out the $10\%$ of genes with lowest mean expression among subjects. We obtain expression values for 985 subjects and $15{,}678$ genes. Subsequently, for each gene we compute a p-value by means of a two-sided two-sample t-test, with the null hypothesis that the gene's mean expression is the same between the populations. The global test statistic for a pathway is defined as the harmonic mean of its genes' p-values \citep{harmonic}. This choice follows the fact that we expect medium-high correlation and quite dense signal, and so we suppose that the harmonic mean will have good power (see Section \ref{sims}).

We consider $\tfrom=\alpha=0.05$ and $\tto=0.5$, and we use the ‘quick' setting described in Section \ref{appl:fmri}. When considering all genes together, $33.08\%$ result to be active. Subsequently, we analyse the 352 pathways contained in the KEGG database \citep{kegg}. The procedure finds 344 active pathways, 56 of which have TDPs higher than the whole gene set. This suggests that differences in gene expression appear mainly in these 56 pathways. Table \ref{Table:gene} contains results for the whole set of genes and the 10 pathways having the highest proportions of active genes: name, function, size and lower confidence bound for the TDP. Applying the procedure to the entire set of genes requires around one minute, while studying all 352 pathways requires around 13 minutes.

\begin{table}
\caption{\label{Table:gene} Differential gene expression data: analysis of pathways. Only the 10 pathways with highest TDP are shown.}
\centering
\begin{adjustbox}{max width=\textwidth}
\begin{tabular}{rrcc}
\toprule
pathway & function & size & TDP ($\%$) \\
$S$ &  & $s$ & $\dsa/s$ ($\%$) \\ 
\midrule
\rowcolor{lightgray} all genes &  & $15{,}678$ & $33.08$ \\
\rowcolor{white} hsa03450 & non-homologous end-joining & 11 & 72.73\\
\rowcolor{white} hsa03050 & proteasome & 44 & 70.45\\
\rowcolor{white} hsa04110 & cell cycle & 120 & 55.83\\
\rowcolor{white} hsa03030 & DNA replication & 36 & 55.56\\
\rowcolor{white} hsa03013 & nucleocytoplasmic transport & 100 & 52.00\\
\rowcolor{white} hsa00900 & terpenoid backbone biosynthesis & 22 & 50.00\\
\rowcolor{white} hsa03267 & virion & 4 & 50.00\\
\rowcolor{white} hsa03008 & ribosome biogenesis in Eukaryotes & 69 & 49.28\\
\rowcolor{white} hsa01210 & oxocarboxylic acid metabolism & 17 & 47.06\\
\rowcolor{white} hsa00450 & selenocompound metabolism & 13 & 46.15\\
\bottomrule
\end{tabular}
\end{adjustbox}
\end{table}

% --------------------------------------------------------------------------------------------------------------------------------------------------------------------------------------------------------------------------------------------------------------------------------------------------------------------
% --------------------------------------------------------------------------------------------------------------------------------------------------------------------------------------------------------------------------------------------------------------------------------------------------------------------

\section{Proofs}\label{proofs}

% --------------------------------------------------------------------------------------------------------------------------------------------------------------------------------------------------------------------------------------------------------------------------------------------------------------------
% SECTION 3

\subsection*{Lemma \ref{L:perm}}
\textit{Under Assumption \ref{A:perm}, the test that rejects $H_S$ when $t_S > t_S^{(\lceil (1-\alpha) B\rceil)}$ is an $\alpha$-level test.}

\begin{proof}
Proof of Lemma \ref{L:perm} is in \citet{exact0} (see Theorem 1).
\end{proof}

% --------------------------------------------------------------------------------------------------------------------------------------------------------------------------------------------------------------------------------------------------------------------------------------------------------------------

\subsection*{Theorem \ref{T:perm}}
\textit{Under Assumption \ref{A:perm}, the test that rejects $H_S$ when $c_S^{(\lfloor \alpha B\rfloor +1)}>0$ is an $\alpha$-level test.}

\begin{proof}
Let $\omega_0=\lceil (1-\alpha)B\rceil$ and $\omega=\lfloor \alpha B\rfloor +1$. Consider the sorted values $c_S^{(1)}\leq\ldots\leq c_S^{(B)}$. Since $c_S^\pi=t_S - t_S^\pi$ for each $\pi$,
\[c_S^{(k)}=t_S - t_S^{(B-k+1)}\qquad (k\in\{1,\ldots,B\}).\]
In particular, for $k=\omega$ we obtain $c_S^{(\omega)}=t_S - t_S^{(\omega_0)}$. Then
\[t_S > t_S^{(\omega_0)}\quad\text{if and only if}\quad c_S^{(\omega)}=t_S - t_S^{(\omega_0)} > 0.\]
By Lemma \ref{L:perm}, under Assumption \ref{A:perm} the test that rejects $H_S$ when $c_S^{(\omega)}>0$ is an $\alpha$-level test.
\end{proof}

% --------------------------------------------------------------------------------------------------------------------------------------------------------------------------------------------------------------------------------------------------------------------------------------------------------------------
% SECTION 5

\subsection*{Lemma \ref{L:phi_q}}
\textit{$\phi(0)=0$ and $\phi(s+1)=1$. Moreover, $\pz=0$ if and only if $z\in\{0,\ldots,q\}$.}

\begin{proof}
When $z=0$, we have $\mathcal{V}_0=2^M$ and $\emptyset\in \mathcal{V}_0\setminus\R$, so $\phi(0)=0$. When $z=s+1$, we have $\mathcal{V}_{s+1}=\emptyset\subseteq\R$, therefore $\phi(s+1)=1$.

Fix a generic $z\in\{0,\ldots,s+1\}$. By definition (\ref{def:pz}), $\pz=0$ if and only if there exists a set $V\subseteq M$ such that $|V\cap S|\geq z$ and $V\notin\R$. By definition (\ref{def:q_first}) of $q$, this is true if and only if $z\leq q$.
\end{proof}

% --------------------------------------------------------------------------------------------------------------------------------------------------------------------------------------------------------------------------------------------------------------------------------------------------------------------

\subsection*{Lemma \ref{L:upper}}
\textit{$\upv\leq\qua$ for all $V\in\Vv$. Hence $\min_v\upv >0$ implies $\pz=1$.}

\begin{proof}
Fix a set $V\in\Vv$, so that $|V|=v$ and $|V\cap S|\geq z$, and a transformation $\pi$. The corresponding centered statistic may be written as
\[c_V^\pi= \sum_{h=1}^{z} c_{\hi_h}^{\pi} + \sum_{h=1}^{v-z} c_{\hj_h}^{\pi}\]
where, similarly to definition (\ref{def:Uz}) of $b^\pi_v$, we have
\begin{align*}
V\cap S=\{\hi_1(\pi),\ldots,\hi_{|V\cap S|}(\pi)\}\quad :\quad &c_{\hi_1(\pi)}^{\pi}\leq\ldots\leq c_{\hi_{|V\cap S|}(\pi)}^{\pi}\\
V\setminus \{\hi_1(\pi),\ldots,\hi_z(\pi)\}
=\{\hj_1(\pi),\ldots,\hj_{v-z}(\pi)\}
\quad :\quad &c_{\hj_1(\pi)}^{\pi}\leq\ldots\leq c_{\hj_{v-z}(\pi)}^{\pi}.
\end{align*}
We compare the definitions of $b_v^{\pi}$ and $c_V^\pi$, starting with the elements in $S$. Since $(V\cap S)\subseteq S$, the statistics in $V\cap S$ cannot be smaller than the first smallest statistics in $S$, i.e.,
\begin{align}
c_{\hi_{h}(\pi)}^{\pi}\geq c_{i_h(\pi)}^{\pi}\quad (h\in\{1,\ldots, |V\cap S|\}). \label{res:i}
\end{align}
A similar comparison can be made for the elements outside $S$. Write
\begin{align*}
\{\hj_1(\pi),\ldots,\hj_{v-z}(\pi)\}&=\{\hi_{z+1}(\pi),\ldots,\hi_{|V\cap S|}(\pi)\}\cup (V\setminus S)\\
\{j_1(\pi),\ldots,j_{m-z}(\pi)\}&=\{i_{z+1}(\pi),\ldots,i_s(\pi)\}\cup (M\setminus S).
\end{align*}
As $(V\setminus S)\subseteq (M\setminus S)$, the statistics in $V\setminus S$ cannot be smaller than the first smallest statistics in $M\setminus S$. By combining this result with (\ref{res:i}), we obtain
\[c_{\hj_{h}(\pi)}^{\pi}\geq c_{j_h(\pi)}^{\pi}\quad (h\in\{1,\ldots, v-z\}).\]
As a consequence, $c_V^\pi\geq b_v^\pi$. Since the inequality holds for any transformation $\pi\in\pib$, it holds also for the quantile, so that $\qua\geq b_v^{(\omega)}= \upv$.

If $\upv > 0$, then $\qua>0$ for all $V\in \Vv$, and so $\Vv\subseteq\R$. Finally, if $\min_v\upv >0$, then $\Vv\subseteq\R$ for each $v\in\{z,\ldots,m\}$, and so $\V\subseteq R$; by definition, $\pz=1$.
\end{proof}

% --------------------------------------------------------------------------------------------------------------------------------------------------------------------------------------------------------------------------------------------------------------------------------------------------------------------

\subsection*{Proposition \ref{P:short}}
\textit{As $\underpz\leq\pz$ for each $z\in\{0,\ldots,s+1\}$, $q^{(0)}\geq q$.}

\begin{proof}
Fix a value $z\in\{0,\ldots,s+1\}$, and suppose that $\underpz=1$. By definitions (\ref{def:upz}) and (\ref{def:upz_bis}), this means that there exists $z^*\in\{0,\ldots, z\}$ with $\min_v\up[z^*]>0$. By Lemma \ref{L:upper}, this implies that $\phi(z^*)=1$ and, since $\phi$ is increasing, $\phi(z)=1$. As $\underpz=1$ implies $\phi(z)=1$, we have $\underpz\leq\pz$. When comparing the change points of $\underline{\phi}$ and $\phi$, we obtain $q^{(0)}\geq q$.
\end{proof}

% --------------------------------------------------------------------------------------------------------------------------------------------------------------------------------------------------------------------------------------------------------------------------------------------------------------------

\subsection*{Theorem \ref{T:d0}}
\textit{$d^{(0)}\leq d$.}

\begin{proof}
From Proposition \ref{P:short} we have $q^{(0)}\geq q$. Since $d^{(0)}=s-q^{(0)}$ and $d=s-q$, we obtain $d^{(0)}\leq d$.
\end{proof}

% --------------------------------------------------------------------------------------------------------------------------------------------------------------------------------------------------------------------------------------------------------------------------------------------------------------------
% SECTION 6

\subsection*{Lemma \ref{L:lower}}
\textit{$\min_v\lowv \leq 0$ implies $\pz=0$.}

\begin{proof}
Suppose that $\min_v\lowv\leq 0$. This means that there exists $v\in\{z,\ldots,m\}$ with $\lowv\leq 0$. Since $\lowv=c_{V_{v}}^{(\omega)}$ with $V_{v}\in\Vv$, we have that $\Vv\not\subseteq\R$. Hence $\V\not\subseteq\R$, and so $\pz=0$.
\end{proof}

% --------------------------------------------------------------------------------------------------------------------------------------------------------------------------------------------------------------------------------------------------------------------------------------------------------------------

\subsection*{Proposition \ref{P:short2}}
\textit{$\underpz\leq\pz\leq\overpz$ for each $z\in\{0,\ldots,s+1\}$. Hence $\underpz=\overpz$ implies $\underpz=\pz$, i.e., equivalence between the shortcut and closed testing.}

\begin{proof}
Fix a value $z\in\{0,\ldots,s+1\}$, and suppose that $\overpz=0$. By definitions (\ref{def:opz}) and (\ref{def:opz_bis}), this means that there exists $z^*\in\{z,\ldots, s+1\}$ with $\min_v\low[z^*]\leq 0$. By Lemma \ref{L:upper}, this implies that $\phi(z^*)=0$ and, since $\phi$ is increasing, $\phi(z)=0$. From the result of Proposition \ref{P:short}, and since $\overpz=0$ implies $\phi(z)=0$, we have $\underpz\leq\pz\leq\overpz$. As a consequence, if $\underpz=\overpz$, then $\underpz=\pz=\overpz$.
\end{proof}

% --------------------------------------------------------------------------------------------------------------------------------------------------------------------------------------------------------------------------------------------------------------------------------------------------------------------

% SECTION 7

\subsection*{Proposition \ref{P:shortiter}}
\textit{For any $n\in\mathbb{N}$ and any $z\in\{0,\ldots,s+1\}$,
\[\underpzn\leq\underpzn[n+1]\leq\underpzn[m]=\pz=\overpzn[m]\leq\overpzn[n+1]\leq\overpzn .\]
Hence $\underpzn=\overpzn$ implies $\underpzn=\pz$, i.e., equivalence between the iterative shortcut and closed testing. Moreover, $q^{(n)}\geq q^{(n+1)}\geq q^{(m)}=q$.}

\begin{proof}
Fix $n\in\mathbb{N}$ and $z\in\{0,\ldots,s+1\}$. First, we prove that $\underpzn\leq\pz\leq\overpzn$; as a consequence, $\underpzn=\overpzn$ implies $\underpzn=\pz$. When we apply the shortcut within a subspace $\V^k$ of $\V$, by Proposition \ref{P:short2} we obtain $\underpz\leq\pz\leq\overpz$. Since this property holds for any subspace, it holds also when we take the minimum of $\underpz$ and $\overpz$ over all subspaces, and so $\underpzn\leq\pz\leq\overpzn$.

Subsequently, we prove that
\[\underpzn\leq\underpzn[n+1]\leq\pz\leq\overpzn[n+1]\leq\overpz ,\]
and so $q^{(n)}\geq q^{(n+1)}\geq q$. The single-step shortcut examines $\V$. If it determines that $\underpz=\pz=\overpz$, the procedure does not partition $\V$, and trivially we have $\underpzn=\pz=\overpzn$ for any $n\in\mathbb{N}$. Otherwise, if $\underpz=0<\overpz=1$, at step $n=1$ we partition $\V$. By Proposition \ref{P:shortiter}, and since $\underline{\phi}^{(1)}$ and $\overline{\phi}^{(1)}$ take values in $\{0,1\}$, we must have $\underpz\leq\underpzn[1]\leq\pz\leq\overpzn[1]\leq\overpz$. The same argument may be applied for any subsequent step $n\in\mathbb{N}$.

Finally we prove that $\underpzn[m]=\pz=\overpzn[m]$, and thus $q^{(m)}=q$. When a subspace contains a single set $V$, both the bound (\ref{def:Uz}) and the path (\ref{def:Lz}) coincide with its quantile $\qua$, and so the shortcut in the subspace must be equivalent to closed testing. Assume the worst case, where $S=M$ and $z=1$, and where the shortcut is not equivalent to closed testing in any subspace containing more than one set. The space of interest is $\mathcal{V}_1=2^M\setminus\{\emptyset\}$, with size $|\mathcal{V}_1|=2^{m}-1$. After $m$ steps, the procedure generates $2^{m}-1$ subspaces, each of them containing exactly one set, and so the shortcut is equivalent to closed testing within each one. As a consequence, $\underpzn[m]=\pz=\overpzn[m]$.
\end{proof}

% --------------------------------------------------------------------------------------------------------------------------------------------------------------------------------------------------------------------------------------------------------------------------------------------------------------------

\subsection*{Theorem \ref{T:dn}}
\textit{$d^{(n)}\leq d^{(n+1)}\leq d^{(m)}=d$ for each $n\in\mathbb{N}$.}

\begin{proof}
From Proposition \ref{P:shortiter}, for each $n\in\mathbb{N}$ we have $q^{(n)}\geq q^{(n+1)}\geq q^{(m)}=q$, and so $d^{(n)}\leq d^{(n+1)}\leq d^{(m)}=d$.
\end{proof}

% --------------------------------------------------------------------------------------------------------------------------------------------------------------------------------------------------------------------------------------------------------------------------------------------------------------------
% SECTION 9

\subsection*{Proposition \ref{P:trunc}}
\textit{Let $V\subseteq M$ and $i\in M$. If $i$ satisfies condition (\ref{condtrunc1}), then $V\in\R$ implies $(V\cup\{i\})\in\R$. If $i$ satisfies condition (\ref{condtrunc2}), then $(V\cup\{i\})\in\R$ implies $V\in\R$.}

\begin{proof}
Recall that $V\subseteq \R$ if and only if $t_V > t_{V}^{(\omega_0)}$ with $\omega_0=\lceil (1-\alpha)B\rceil$ (equivalence between Lemma \ref{L:perm} and Theorem \ref{T:perm}). Fix an index $i\in M$, and define $Q=V\cup\{i\}$. If $i\in V$, we obtain the trivial case where $Q=V$, hence suppose that $i\in M\setminus V$. In this case, since $\tto\leq\tfrom$, for each $\pi\in\pib$ we have
\begin{align*}
f(t_i^\pi)=\tto\cdot\mathbf{1}\{t_i^\pi<\tfrom\} + t_i^\pi\cdot\mathbf{1}\{t_i^\pi\geq\tfrom\}\geq \tto\\
t_Q^\pi = \sum_{i\in Q}f(t_i^\pi)=t_V^\pi + f(t_i^\pi).
\end{align*}

First, assume that $V\in\R$ and that property (\ref{condtrunc1}) holds:
\begin{align*}
f(t_i)\geq \tto\quad&\Longrightarrow\quad t_Q\geq t_V + \tto\\
f(t_i^\pi)= \tto\quad&\Longrightarrow\quad t_Q^\pi= t_V^\pi + \tto\qquad (\pi\neq\text{id}).
\end{align*}
All test statistics for $Q$ coincide to those for $V$ plus a constant $\tto$, with the exception of the observed statistic. Since $V\in\R$, $t_V > t_V^{(\omega_0)}$, and so when ordering the statistics we obtain $t_Q^{(k)} = t_V^{(k)} +\tto$ for all $k\leq\omega_0$. Therefore
\[t_Q\geq t_V +\tto > t_V^{(\omega_0)}+\tto = t_Q^{(\omega_0)}\]
and thus $Q\in\R$.

Subsequently, assume that $Q\in\R$ and that property (\ref{condtrunc2}) holds:
\begin{align*}
f(t_i)= \tto\quad&\Longrightarrow\quad t_Q= t_V + \tto\\
f(t_i^\pi)\geq \tto\quad&\Longrightarrow\quad t_Q^\pi\geq t_V^\pi + \tto\qquad (\pi\neq\text{id}).
\end{align*}
Therefore $t_Q^{(\omega_0)} \geq t_V^{(\omega_0)} + \tto$ and
\[t_V=t_Q - \tto > t_Q^{(\omega_0)} - \tto \geq t_V^{(\omega_0)}, \]
and thus $V\in\R$.
\end{proof}

% --------------------------------------------------------------------------------------------------------------------------------------------------------------------------------------------------------------------------------------------------------------------------------------------------------------------

% --------------------------------------------------------------------------------------------------------------------------------------------------------------------------------------------------------------------------------------------------------------------------------------------------------------------

% APPENDIX A

\subsection*{Lemma \ref{L:complexity_sumSome1}}
\textit{In the worst case, the computational complexity of Algorithm \ref{algorithm:short} is of order $mB\log(mB)$.}

\begin{proof}
To compute the values of the bound $\upv$ for $v\in\{z,\ldots,m\}$ as in \eqref{def:Uz}, the algorithm operates as following. First, it sorts the centered test statistics for each permutation as in \eqref{permbound1} and \eqref{permbound2}, with a number of operations of order $B\{s\log s + (m-z)\log(m-z)\}$. Subsequently, it computes
\begin{align*}
\upv[z]=b_z^{(\omega)}\quad&\text{where}\quad b_z^{\pi} = \sum_{h=1}^{z} c_{i_h}^{\pi}\quad (\pi\in\pib)\\
\upv=b_{v}^{(\omega)}\quad&\text{where}\quad b_v^{\pi} = b_{v-1}^{\pi} + c_{j_{v-z}}^{\pi}\quad (v\in\{z+1,\ldots,m\},\;\pi\in\pib).
\end{align*}
This requires $mB$ sums and $m-z+1$ sortings of $B$ elements, and so it uses $mB + (m-z+1)B\log B$ operations.

In the worst case, when $s=m$ and $z=1$, the total number of operations needed to compute $\up$ is of order $mB\{\log m +\log B\}=mB\log(mB)$. The same argument applies to the path $\low$.
\end{proof}

% --------------------------------------------------------------------------------------------------------------------------------------------------------------------------------------------------------------------------------------------------------------------------------------------------------------------

\subsection*{Lemma \ref{L:complexity_sumSome1bis}}
\textit{In the worst case, Algorithm \ref{algorithm:bab} converges after a number of iterations of order $2^m$, where each iteration has complexity of order $mB\log(mB)$.}

\begin{proof}
In the worst case, $s=m$, $z=1$, and all subspaces containing more than one set need to be partitioned. After $m$ steps, the procedure generates $2^{m}-1$ subspaces, each of them containing exactly one set. In this case, the total number of iterations of the single-step shortcut is of order $2^m$. The computational complexity of each iteration is given by Lemma \ref{L:complexity_sumSome1}.
\end{proof}

% --------------------------------------------------------------------------------------------------------------------------------------------------------------------------------------------------------------------------------------------------------------------------------------------------------------------

\subsection*{Lemma \ref{L:complexity_sumSome2}}
\textit{In the worst case, embedding Algorithm \ref{algorithm:short} into a binary search requires a number of operations of order $mB(\log^2 m +\log B)$.}

\begin{proof}
By Lemma \ref{L:complexity_sumSome1}, the single-step shortcut of Algorithm \ref{algorithm:short} has complexity at most of order $mB\log(mB)$. In the worst case, when $s=m$, the binary search applies the shortcut at most $\log_2 m$ times \citep{knuth}. Hence the worst-case complexity is of order $mB\log(mB)\log m=mB(\log^2 m +\log B)$.
\end{proof}

% --------------------------------------------------------------------------------------------------------------------------------------------------------------------------------------------------------------------------------------------------------------------------------------------------------------------

\subsection*{Lemma \ref{L:upper_shape}}
\textit{Define $\ci[1]$ as in (\ref{def:c1}), and $\ci[2]$ as in (\ref{def:c2}). Then $\up(z)\geq\up(z+1)\geq\ldots\geq\up(\ci[1])$, and $\up(\ci[2])\leq \up(\ci[2] + 1)\ldots\leq\up(m)$.}

\begin{proof}
For any $v\in\{z+1,\ldots,m\}$, we have
\[\upv[v-1]=b_{v-1}^{(\omega)}\quad\text{where}\quad b_{v-1}^\pi=b_{v}^\pi - c_{j_{v-z}}^\pi\quad (\pi\in\pib).\]
If $v\leq \ci[1]$, then $c_{j_{v-z}(\pi)}^\pi\leq 0$ for all $\pi$, and so $\up(v-1)\geq\upv$; as a consequence, $\upv[z]\geq\upv[z+1]\ldots\geq \upv[\ci[1]]$. Similarly, if $v > \ci[2]$, then $c_{j_{v-z}(\pi)}^\pi\geq 0$ for all $\pi$, and so $\up(v-1)\leq\upv$; therefore $\upv[\ci[2]]\leq\upv[\ci[2]+1]\leq\ldots\leq\upv[m]$.
\end{proof}

% --------------------------------------------------------------------------------------------------------------------------------------------------------------------------------------------------------------------------------------------------------------------------------------------------------------------

\subsection*{Lemma \ref{L:b_in_sub}}
\textit{$\lowv$ is the same in $\V$ and $\Vrem$ for each $v\in\{z,\ldots,m-1\}$.}

\begin{proof}
By definition (\ref{def:Lz}), in $\V$ the path is $\lowv=\qua[V_v]$ with $V_v=\{i_1,\ldots,i_z\}\cup\{j_1,\ldots,j_{v-z}\}$ for any $v\in\{z,\ldots,m\}$. Recall that $\Vrem=\{V\in\V\,:\,j^*\notin V\}$, where $j^*=j_{m-z}$ (see Section \ref{babintr}). Therefore in $\Vrem$ the greatest set is $M\setminus\{j^*\}$, and so the path $\low$ is defined for sizes $v\in\{z,\ldots,m-1\}$. Moreover, $j^*\notin V_v$ for all $v\in\{z,\ldots,m-1\}$, hence in $\Vrem$ the path is defined as in $\V$.
\end{proof}

\end{document}